\documentclass{article}[12pt]
\usepackage{amsmath,amssymb,amsfonts}
\usepackage{graphics, subfigure,color} 

\usepackage{latexsym}
\usepackage[dvips]{graphicx}
\usepackage{epsfig}

\usepackage{hyperref}

\numberwithin{equation}{section}

\newtheorem{prop}{Proposition}[section]

\newtheorem{lemma}[prop]{Lemma}

\newtheorem{theorem}[prop]{Theorem}

\newcommand{\bQ}{{\bf Q}}
\newcommand{\bV}{{\bf V}}
\newcommand{\bX}{{\bf X}}
\newcommand{\bA}{{\bf A}}

\newcommand{\bP}{{\bf P}}

\newcommand{\cI}{{\cal{I}}}

\newcommand{\cG}{{\cal{G}}}

\newcommand{\cF}{{\cal{F}}}

\newcommand{\cB}{{\cal B}}
\newcommand{\cT}{{\cal T}}
\newcommand{\cS}{{\cal S}}
\newcommand{\cJ}{{\cal J}}
\newcommand{\cL}{{\cal L}}

\newcommand{\cM}{{\cal M}}

\newcommand{\cC}{{\cal C}}
\newcommand{\cO}{{\cal O}}
\newcommand{\cW}{{\cal W}}

\newcommand{\Tr}[1]{\:{\rm Tr}\,#1}
 
\newcommand{\bbone}{{\bf 1}}

\newcommand{\be}{\begin{equation}}
\newcommand{\bee}{\begin{equation}}
\newcommand{\ee}{\end{equation}}
\newcommand{\beq}{\begin{eqnarray}}
\newcommand{\eeq}{\end{eqnarray}}
\newcommand{\bqa}{\begin{eqnarray}}
\newcommand{\eqa}{\end{eqnarray}}
\newcommand{\bea}{\begin{eqnarray}}
\newcommand{\eea}{\end{eqnarray}}
\newcommand{\beann}{\begin{eqnarray*}}
\newcommand{\eeann}{\end{eqnarray*}}

\newcommand{\prf}{{\noindent \bf Proof\; \; }}
\newcommand{\qed}{{\hfill $\Box$}}

\makeatletter

\begin{document}

 \title{Corrected  Loop Vertex Expansion
for $\Phi^{4}_2$ Theory}
\author{Vincent Rivasseau,\  Zhituo Wang\\
Laboratoire de Physique Th\'eorique, CNRS UMR 8627,\\
Universit\'e Paris XI,  F-91405 Orsay Cedex, France\footnote{E-mail: rivass@th.u-psud.fr,  zhituo.wang@th.u-psud.fr}
}

\maketitle
\begin{abstract}
This paper is an extended erratum to \cite{RZW}, in which the classic construction and Borel summability of the
$\phi^4_2$ Euclidean quantum field theory was revisited combining a multi-scale analysis with the constructive method called Loop Vertex Expansion (LVE).
Unfortunately we discovered an important error in the method of \cite{RZW}. 
We explain the mistake, and provide a new, correct construction of the $\phi^4_2$
theory according to the LVE. 
 \end{abstract}

\section{Introduction}
\label{intro}

The Loop Vertex Expansion (LVE) \cite{Rivasseau:2007fr, MR1, RW} is a constructive field theory technique
which combines a forest formula \cite{BK, AR1}, a replica trick and the intermediate field 
representation. It can compute the connected functions of stable Bosonic quantum field theories 
with both infrared and ultraviolet cutoffs and a small-coupling without introducing the 
space-time lattices and cluster expansions of standard constructive theory
\cite{GJS,GJ,Riv}. It can also be considered as a convergent reshuffling of Feynman graphs
using canonical combinatorial tools \cite{Rivasseau:2013ova}.

To remove cutoffs in models with ultraviolet divergences requires to extend the LVE technique to allow for renormalization,
that is for the explicit cancellation of counterterms. A first method, called ``cleaning expansion"
has been introduced to cure the ultraviolet divergences of superrenormalizable 
models such as the commutative $\phi^4_2$ model \cite{RZW} and the 
2-dimensional Grosse-Wulkenhaar model \cite{ZTW}. 

We recently discovered an important error in this method and in particular in the bounds of \cite{RZW}. 
Unfortunately it does not seem possible to fix the cleaning expansion itself. Since
it expands all loop vertices up to infinity, an arbitrary number of them can require renormalization, and
the expansion fails to converge. The attentive reader can trace back
the mistake to the section 5 of \cite{RZW}, where some crucial sums over the
scales of the tadpoles were simply forgotten. 

In this paper we correct this mistake, replacing the ``cleaning expansion" with a more careful expansion, which keeps hardcore constraints 
between the exponential of the interaction in different slices. This multiscale loop vertex expansion has been 
defined and checked to work in detail for a simpler toy model in \cite{MLVE}. Here we adapt it to the $\phi^4_2$ 
theory, restricting ourselves for simplicity to remove the ultraviolet cutoff only. Hence we keep the theory in a single unit square.

Of course the construction of $\phi^4_2$ \cite{GJS,nelson, Simon} and its Borel summability \cite{EMS} are classic milestones of constructive theory. 
The purpose of this paper is just to obtain these results again with a multislice LVE. We prove analyticity in the coupling constant of the free energy\footnote{Extensions to arbitrary Schwinger functions are left as an exercise to the reader.} 
of the theory in a cardioid-shaped domain (see Figure \ref{cardio})
which has opening angle arbitrarily close to $2 \pi$. This domain, which is the natural one for LVE-type expansions \cite{Gurau:2013pca}, is larger 
than the ones usually considered in the constructive literature \cite{Riv,EMS}, which are of Nevanlinna-Sokal or Watson-type, 
hence have opening angles respectively $\pi $ or $\pi + \epsilon$.

It is also our hope and goal to adapt ultimately the MLVE techniques to treat just-renormalizable models.
An important motivation for this is the constructive treatment of quantum field theories with non-local interactions such as the four dimensional Grosse-Wulkenhaar non-commutative field theory \cite{Grosse:2004yu}
or tensor group field theories \cite{ BenGeloun:2011rc,Carrozza:2013wda}, for which the LVE seems clearly better adapted than traditional constructive techniques
\cite{Gurau:2013pca,sefu1,Delepouve:2014bma}.

We refer to \cite{Brydges:2014nba,Brydges:2014qba,Bauerschmidt:2014pba,Brydges:2014rba,Brydges:2014sba} for a recent extensive reformulation of the constructive
renormalization group multiscale techniques
relying on scaled lattices.

\section{The Model and its Slice-Testing Expansion}

We consider the free Bosonic $\phi^4_2$ theory in a fixed volume, namely the unit square $[0,1]^2$ with coupling constant $\lambda$. From now on any spatial 
integral has to be understood as restricted to $[0,1]^2$ and we use the notation $\Tr$ to mean $\int_{[0,1]^2} d^2x$.
The formal partition function of the theory with source $J$ is
\begin{equation}\label{model}
Z(J,\lambda)=\int d\mu_{C} ( \phi ) e^{\Tr J(x) \phi (x) 
-\frac{\lambda}{2}\Tr\; \phi^4(x)}
\end{equation}
where $d\mu_C$ is the normalized Gaussian measure with 
covariance or propagator 
\begin{equation}
C(x, y)=\frac{1}{4\pi}\int_0^\infty\frac{d\alpha}{\alpha}e^{-\alpha
m^2-\frac{(x-y)^2}{4\alpha}},
\end{equation}
with $x$ and $y$ restricted to $[0,1]^2$, hence with free boundary conditions; the theory could equally well be considered
on the two dimensional torus with periodic boundary conditions without significant change in the analysis. 
A main problem in quantum field theory is to compute the logarithm $\log Z (J,\lambda)$ which is the generating function 
of the connected Schwinger functions. 

The covariance at coinciding points $ C(x, x) = T$ corresponds to a self-loop or
tadpole in perturbation theory. It diverges logarithmically in the ultraviolet cutoff and
is the only primitive ultraviolet divergence of the theory. Renormalization reduces to Wick ordering, hence
the  renormalized model 
has partition function:
\bea
Z(J,\lambda)&=&\int d\mu_{C} ( \phi ) e^{\Tr J(x) \phi (x) 
-\frac{\lambda}{2}\Tr\; :\phi^4(x):} = \int d\mu_{C} e^{\Tr J(x) \phi (x) -\frac{\lambda}{2}\Tr \; [\phi^4-6T \phi^2 + 3T^2]}\nonumber \\
&=& \int d\mu_{C} e^{\Tr J(x) \phi (x) -\frac{\lambda}{2}\Tr\;   [(\phi^2-3T)^2  -6 T^2]}, \label{model}
\eea
where the Wick ordering in $:\phi^4(x): \equiv \phi^4-6T \phi^2 + 3T^2$ is taken
with respect to $C$ \cite{Simon}.

These expressions are formal and to define the theory one needs to introduce an ultraviolet cutoff.
This is most conveniently done in a multiscale representation \cite{Riv} which
slices the propagator in the parametric representation, then keeps only a finite number of slices.

%<<<<<<<<<<<<<<<<<<<<<<<<<<<<<<<<<<<<<<<<<<<<<<<<<<<<<<<<<<<<<<<<<<<<<<<<<<<<<>>>>>>>>>>>>>>>>>>>>>>>>>>>>>>>>>>>>>>>>>>>>>>>>>>>>>>>>>>>>>
\subsection{Slices and Intermediate Field Representation}

We fix an integer $M>1$,  and we slice the propagator as usual in the multislice analysis \cite{MLVE,Riv}, defining the ultraviolet cutoff as a maximal slice index $j_{max}$. The ultraviolet limit 
corresponds to $j_{max} \to \infty$. Hence we define\footnote{Beware we choose the convention of \emph{lower} indices for slices, as in \cite{MLVE}, not upper
indices as in \cite{Riv}.}
$C = C_{\le j_{max}}  =\sum_{j=0}^{j_{max}} C_j$ with:
\bea
C_0(x, y)&=&\int_{1}^{\infty}e^{-\alpha
m^2-\frac{(x-y)^2}{4\alpha}}\ \frac{d\alpha}{\alpha}\le
Ke^{-c |x-y|}, \\
C_j(x, y)&=&\int_{M^{-2j}}^{M^{-2(j-1)}}e^{-\alpha
m^2-\frac{(x-y)^2}{4\alpha}}\ \frac{d\alpha}{\alpha}\le
O(1)e^{-O(1) M^j|x-y|} \;\; {\rm for}\; j\ge 1. \label{multbound}
\eea
Throughout this paper we use the time-honored constructive convention of noting $O(1)$ any inessential constant. 
The set of slice indices $\cS = [0, j_{max}]$ has $1 + j_{max}$ elements.

We also put $T_j =C_j (x,x)  \le  O(1)$, and
\bea
C_{\le j}    = \sum_{k=0}^{j} C_k  = \int_{M^{-2j}}^{\infty} e^{-\alpha m^2-\frac{(x-y)^2}{4\alpha}}\ \frac{d\alpha}{\alpha}\quad, \quad
T_{\le j} = \sum_{k=0}^{j}  T_k  , \eea

The partition function $Z^{j_{max}}(J, \lambda)$ obtained by substituting $C_{\le j_{max}}$ instead of $C$ and $T_{\le j_{max}}$
instead of $T$ in \eqref{model} is now well defined. For simplicity we put from now on the source $J=0$, defining $Z(\lambda) = Z(J,\lambda) \vert_{J=0}$. 
The ultraviolet limit of the free energy 
\bee p (\lambda) = \lim_{j_{max} \to \infty }\log Z^{j_{max}}(\lambda) \ee
exists and it is a Borel summable function of $\lambda$.
This paper is devoted to recover these classical results \cite{nelson,GJS, Simon,EMS} in the LVE representation.
 For simplicity we may continue to write
$C$ and $T$ from now on, in which case they mean $C_{\le j_{max}}$ and $T_{\le j_{max}}$.

The main problem of this model compared to the toy model of \cite{MLVE} is that this action is not positive \cite{nelson}.
Observe that for $\lambda >0$ 
\bee   e^{-\frac{\lambda}{2}\Tr [(\phi^2-3T)^2  -6 T^2]}  \le e^{3 \lambda T^2} , \label{theory1}
\ee
producing Nelson's famous divergent bound \cite{nelson}  as $j_{max} \to \infty$:
\bee  \vert Z^{j_{max}}(\lambda) \vert \le e^{ \lambda O(1) j_{max}^2}   . \label{nelbound}
\ee

\begin{figure}[!t]
\begin{center}
{\includegraphics[width=7cm]{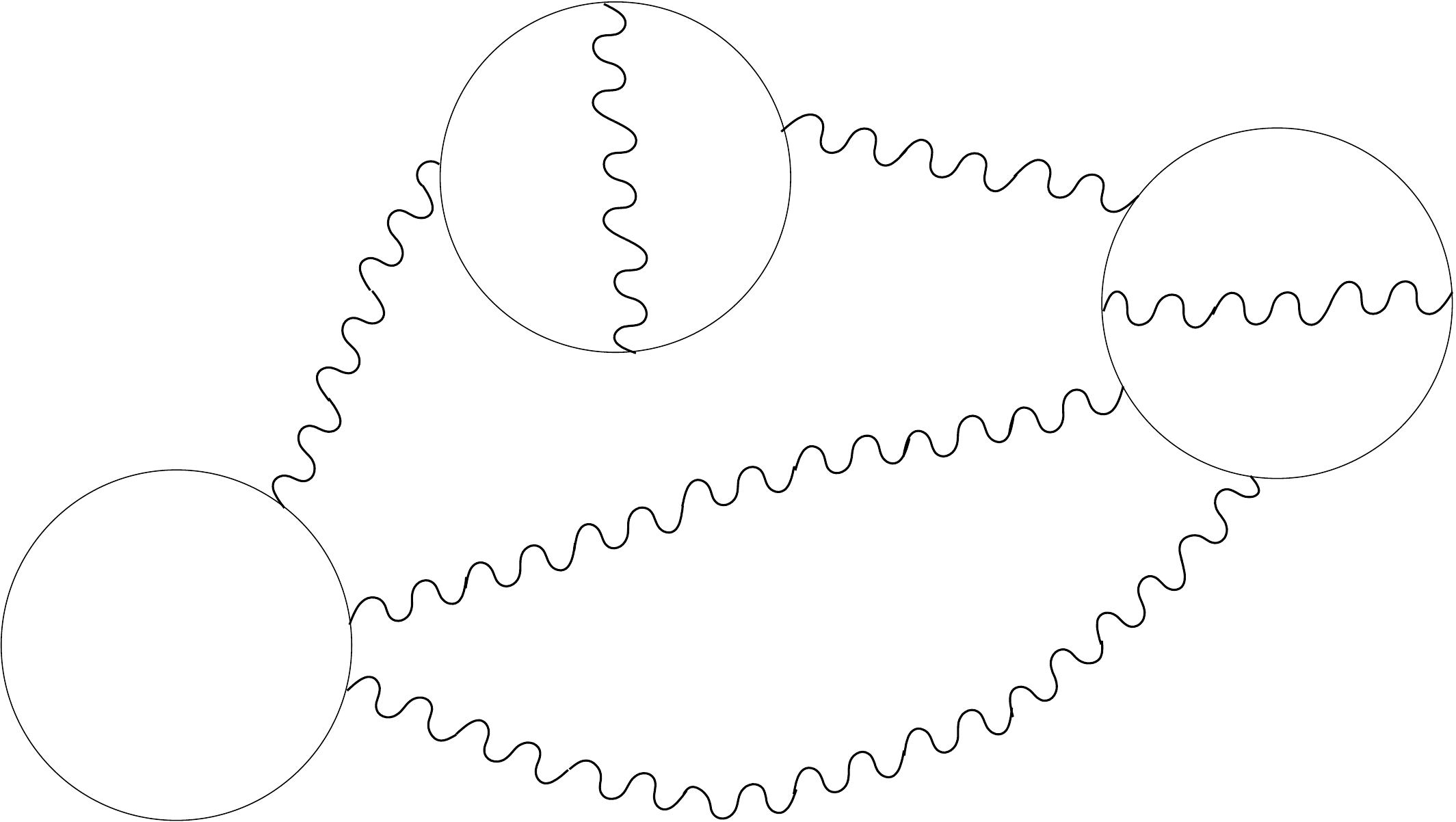}}
\end{center}
\caption{An Intermediate Field Graph}
\label{graphinter6}
\end{figure}

Introducing the intermediate field $\sigma$, integrating out the
terms that are quadratic in $\phi(x)$ and using that $\Tr T\sigma = \Tr C \sigma $, we get
\bea
Z^{j_{max}}(\lambda)&=&\int d\nu(\sigma) e^{\Tr  \bigl(3\lambda  T^2  + 3i \sqrt{\lambda} T
\sigma-\frac{1}{2} \log[1+2i \sqrt{\lambda}C \sigma ]  
\bigr)}, \label{expre0} 
\\ &=&\int d\nu(\sigma) e^{ \Tr  \bigl(3\lambda  T^2  + 2i \sqrt{\lambda} T
\sigma-\frac{1}{2} \log_2[1+2i \sqrt{\lambda}C\sigma ] 
\bigr)}, \label{expre}
\eea
where $d\nu(\sigma)$ is the
ultralocal measure on $\sigma$ with covariance $\delta(x-y)$, and the function
\bee \log_2 (1-x) \equiv x+ \log (1-x) = O(x^2)
\ee 
has to be defined in the operator sense, by the kernel:
\bee [  \log_2 (1+2i \sqrt{\lambda}C\sigma )] (x,y)=- \sum_{k=2}^{\infty} \frac{(-2i \sqrt{\lambda})^k}{k}   \int d^2x_1 \cdots  \int d^2x_{k-1}  \bigl[  C (x,x_1) \sigma (x_1)  
C (x_1,x_2)  \cdots  C (x_{k-1},y) \sigma (y)\bigr]. \label{loops}
\ee
(Adding a trace in the left hand side of \eqref{loops} would corresponds to multiply the right hand side by $\delta(x-y)$ and to integrate over $x$ and $y$.) 
The perturbation theory in terms of $\sigma$ is indexed by intermediate field Feynman graphs (see Figure \ref{graphinter6}) whose vertices are the loops obtained by the expansion \eqref{loops} into traces,
and whose $\sigma$-propagators, represented by wavy lines in Figure \ref{graphinter6}, correspond to the former $\phi^4$ \emph{vertices} of ordinary perturbation expansion, hence bear a coupling constant $\lambda$. The loop vertices are themselves cycles of the old $\phi^4$
propagators, which now occur at each \emph{corner} of the loop vertices\footnote{Remark that such intermediate field
Feynman graphs are really combinatorial maps \cite{Gurau:2013pca,Gurau:2014vwa}. It means that we can define a clockwise
cyclic ordering at each loop vertex. The notion of the \emph{next} intermediate $\sigma$ field (or $\sigma$ half-propagator) at any propagator is then well-defined.}. We 
call these corner $\phi^4$ propagators simply \emph{c-propagators} for short.
The perturbative order of an intermediate graph is the total number of $\sigma$-propagators. In the case of vacuum graphs, which (for simplicity) is the only one 
considered in this paper, it is also half the number of c-propagators. Adding sources would introduce, in addition to the loop vertices, resolvents, which can be considered as ciliated loop vertices
\cite{Gurau:2013pca}. This extension is left to the reader.

\subsection{Flipping Symmetry}

\begin{figure}[!t]
\begin{center}
{\includegraphics[width=7cm]{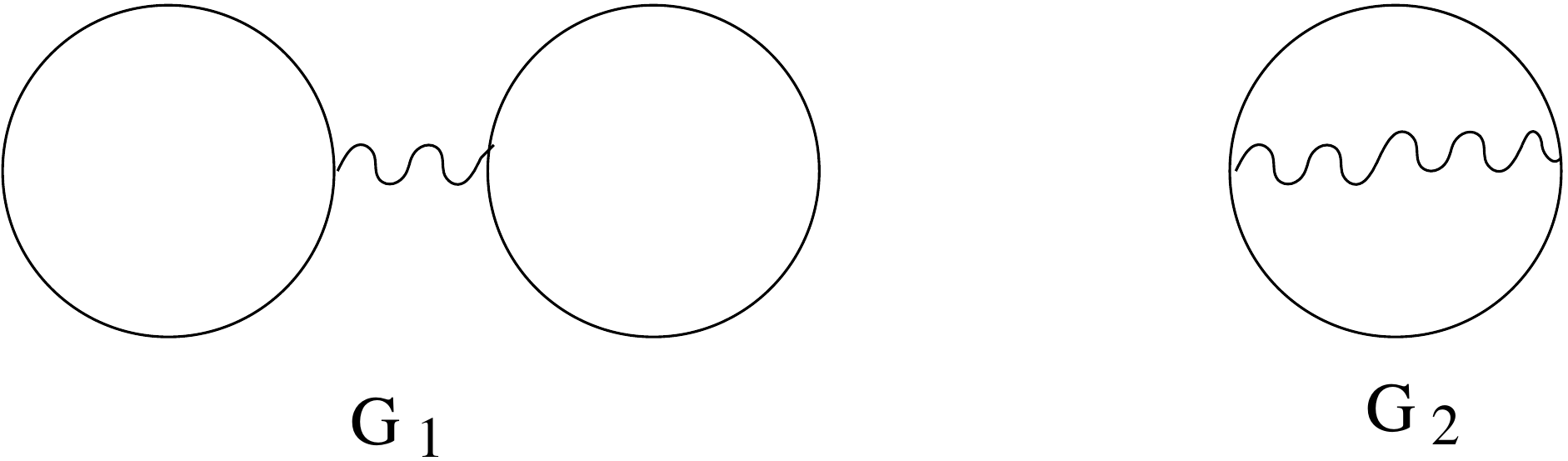}}
\end{center}
\caption{First Order Intermediate Field Graphs}
\label{graphinter7}
\end{figure}

Remark that the initial vertex could have been decomposed in three different ways into two corners joined by a $\sigma$-propagator. It results
in the existence of dualities, which in this context we could simply call \emph{flipping symmetries}: 
any $\sigma$-propagator, being a $\delta$ function, 
can be ``flipped" according to a symmetry group of order 3 by branching the four half-c-propagators into other
pairs. These flips do not change the value of the associated amplitude. It means that many intermediate graphs which do not look the same
have in fact the same amplitude. 
For instance the two first order graphs $G_1$ and $G_2$ of Figure \ref{graphinter7} have the same value, with combinatorial weights respectively 1 and 2, leading to the possibility to 
express the theory at order 1 with the single graph $G_1$ but with combinatorial weight 3 \cite{Rivasseau:2013ova}. 

Since the loop interaction in \eqref{expre} is a $\log_2$, the expansion into intermediate field Feynman graphs has no \emph{loop vertex 
of length one}, i.e. no loop vertex with a single corner-propagator. However renormalization in the intermediate 
field expansion does not reduce to this observation. Wick-ordering puts to zero
all tadpoles in the ordinary perturbative expansion. Accordingly,
the counterterm $2i \sqrt{\lambda} T \sigma$ in \eqref{expre} will precisely 
compensate any intermediate field Feynman graph in which a 
$\sigma$-propagator has  \emph{length one}, that is directly joins the two ends of a c-propagator. Indeed such a $\sigma$ propagator of length one can be flipped
into a loop vertex of length one. This 
is a key difficulty of the LVE in the scalar $\phi^4$ theory, ultimately related to Nelson's bound \eqref{nelbound}.
The intermediate field representation breaks the discrete dualities of the vertex, 
and symmetry breaking, as usual, makes renormalization more difficult \footnote{
This difficulty occurs also in matrix models with quartic interaction, since their vertex has a duality (of order 2 instead of 3).
It does not occur in vector-models \cite{MLVE}, 
nor in tensor models with melonic quartic interactions \cite{DR}, since their vertex has no dualities.}.

This difficulty requires us to perform in the next subsection \ref{slitest} an additional ``slice-testing" expansion, which is the correct substitute
of the ill-fated ``cleaning-expansion" of \cite{RZW}.
For any slice $j$, starting with $j_{max}$ and working up to $j=1$, it searches for the presence of \emph{one}
c-propagator or \emph{one} $2i \sqrt{\lambda} T_j$ counterterm of that slice. Then it performs a Wick contraction  of the $\sigma$ 
field \emph{next} to the c-propagator $C_j$, and also of the $\sigma$ field attached to the $T_j$ counterterm, allowing an explicit compensation between 
tadpole graphs and their counterterms, ultimately bringing convergent factors $M^{-j/2}$ which will tame Nelson's bound.

However there is a difficulty. 
Testing for the presence of a c-propagator $C_j$ with an interpolation parameter $t_j$ and a Taylor expansion step
does not creates just a $C_j$. 
Since $C_j$'s occur within the $\log_2[1+2i \sqrt{\lambda}C\sigma ] $ interaction of \eqref{expre},
derived propagators come equipped with a \emph{resolvent}, which is the operator-valued function of the intermediate field defined by
\bee  R(\sigma) \equiv [1+2i \sqrt{\lambda}C\sigma ]^{-1}.
\ee
The amplitudes of the combinatorial objects obtained at the end of this slice-testing expansion belong therefore to a a new class,
which we call \emph{resolvent amplitudes} and which we now describe.

The resolvent amplitude $A_G$ of an intermediate-field graph $G$ (with slice attributions $\{j(\ell) \}$) is defined as
\bee  A_G (t, \sigma) =  \prod_{v \in V(G)} \bigl[ (-\lambda) \int_{[0,1]^2}  d^2 x_v  \bigr]  \prod_{\ell \in CP(G)} [ R(\sigma)C_{j(\ell)}  ] (x_\ell, x'_\ell ),
\label{ampres}
\ee 
where $V(G)$ is the set of $\sigma$-propagators of $G$,
$CP(G)$ is the set of c-propagators of $G$, and $x_\ell, x'_\ell$ are the positions of the two vertices
at the ends of the c-propagator $\ell$.

The \emph{renormalized} amplitude of the same graph is the same, but with subtracted tadpole resolvents:
\bee  A^R_G (t, \sigma) =  \prod_{v \in V(G)} \bigl[ (-\lambda) \int_{[0,1]^2}  d^2 x_v  \bigr]  \prod_{\genfrac{}{}{0pt}{}{\ell \in CP(G)}{\ell\; tadpole}}  [(R -1)C_{j(\ell)} (\sigma)] (x_\ell, x'_\ell ) \prod_{\genfrac{}{}{0pt}{}{\ell \in CP(G)}{\ell \; not \; tadpole}}  [   R(\sigma)C_{j(\ell)}] (x_\ell, x'_\ell ).
\label{ampresren}
\ee 
Hence in renormalized resolvent amplitudes, c-propagators are not ordinary $C's$ but either
$(R-1)C$ or $RC$ depending whether the c-propagator is a tadpole or not. They correspond therefore, when expanding the $R$
or $R-1$ factors and performing the $\sigma$ Wick contractions, to \emph{infinite series}
of ordinary $\phi^4$ graphs, but with the particularity that the initial propagators $C$'s of the renormalized resolvent graph
cannot be tadpoles. For instance at order 1 there are two resolvent graphs
associated to the two intermediate graphs of Figure \ref{graphinter7}, simply replacing both c-propagators by $(R-1)C$ factors, since they are
both tadpoles. Again these two graphs are equal, allowing to possibly simplify sums over resolvent graphs.

We need two more definitions. Consider any \emph{slice-subset} $J : \{j_1, \cdots j_p\} \subset \cS = [0, \cdots, j_{max}]$. 
A $J$-resolvent graph is defined as a resolvent graph in which exactly
$p$ c-propagators bear marks $\{ j_1, \cdots  , j_p\}$. Finally a 
$J$-resolvent graph is called \emph{minimal} if every connected component of the graph bears at least a mark,
and the total perturbative order of the graph, i.e. the  total number of $\sigma$-propagators, is at most $\vert J \vert$. The set of minimal
$J$-resolvent graphs is noted $\cG (J)$, and we denote $\cG = \cup_{J} \cG (J)$. By convention we could say that
$J = \emptyset$ is allowed, resulting in a single ``empty graph" in $ \cG (\emptyset)$ with no propagator. It will correspond to the free theory, hence to 
the term 1 in the expansion of $  Z^{j_{max}}(\lambda)$ below.

The slice-testing expansion of the next subsection will test through a Taylor remainder formula the presence of a marked (non-tadpole)
propagator in any slice, hence will result in expressing the partition function of the theory as a sum over graphs in $\cG$ multiplying a remaining interaction.
Then a factorization expansion quite similar to that of \cite{MLVE} can be performed. The marked propagators provide the good factors
which ultimately pay for the Nelson bound and all combinatorics.

\subsection{Slice-testing Expansion} \label{slitest}

We introduce inductively interpolation parameters $t_j \in [0,1]$ for the $j$-th scale of the propagators. 

We shall write 
simply $t$ for the family $\{t_j\}, 0\le j \le j_{max}$. It means that we write
\bee  C(t)  = \sum_{j=0}^{j_{max} } t_j C_j , \quad  T(t)  = \sum_{j=0}^{j_{max} }  t_j T_j ,\;\;  T_j = C_j (x, x),
\ee
\bee V (t)   = \Tr  \bigl( 3\lambda  T^2(t) + 2i \sqrt{\lambda} T(t) 
\sigma-\frac{1}{2} \log_2[1+2i \sqrt{\lambda}C(t)\sigma ]  \bigr) . \label{simpleslic}
\ee

We perform a single first order Taylor expansion step in each $t_j$ between 0 and 1. It results in
\bee  Z^{j_{max}}(\lambda)=   \sum_{J \subset \cS} \int d\nu  (\sigma)  \prod_{j \in J}  \int_0^1 dt_j\frac{d}{dt_j} e^{ V (t) } \vert_{t_j = 0 \; {\rm for} \; j \; \not\in J} . \label{testing}
\ee

Each $\frac{d}{dt_j}$ hits either a propagator or a tadpole, resulting in a well defined $T_j$ or a well-defined $C_j \sigma$ brought down from the exponential.
Using that $(\log_2 (1+x) )' = (1+x)^{-1} -1 = - x/(1+x)$, the $T_j$'s  comes equipped with a $T(t) $ or $ \sqrt{\lambda} \sigma $, and the $C_j$'s
comes equipped with an $R(t)  -1$, where 
\bee  R(t) = [1+2i \sqrt{\lambda}C(t)\sigma ]^{-1} .
\ee

When no confusion can result, we should shorten formulas by writing
$C$, $V$ or $R$ for $C(t) $, $V(t) $ and $R(t)$.

We now explicit the result of computing \eqref{testing} at first and second order (i.e. for $\vert J\vert =1 $ and  $\vert J\vert =2$), before
describing the general case through Lemma \ref{testinglemma}.

Applying a first derivative term $\frac{d}{dt_{j_1}} $ in \eqref{testing} gives
\bee 
I_1 = \int \frac{d}{dt_{j_1}}   e^{ V} d\nu  (\sigma) =  \int e^{ V} d\nu  (\sigma) \Tr  \bigl[ 6 \lambda T_{j_1}    T+ 2i \sqrt{\lambda} T_{j_1} \sigma 
- i \sqrt{\lambda}  C_{j_{1}} \sigma (R -1)  \bigr] . \label{deriv1}
\ee
To simplify the expression and explicit the cancellation with the $T_j$ counter term we need to sandwich an  integration by parts
of the two $\sigma$ fields hooked to $C_{j_{1}}$ and $T_{j_{1}}$ between the derivatives
computations. Of course it does not change the value of the result, but simplifies its writing.

Contracting the two explicit $\sigma$ fields in the right hand side of \eqref{deriv1} gives
\bea  
I_1&=&\int e^{ V } d\nu  (\sigma) \Tr \bigl[ 6 \lambda T_{j_1} T + 2i \sqrt{\lambda} T_{j_1} \sigma 
- i \sqrt{\lambda}   C_{j_{1}} \sigma ( R -1)   \bigr] \nonumber \\
&=&
\int e^{ V }  d\nu  (\sigma) \biggl[ 6 \lambda T_{j_1} T    - 4 \lambda T_{j_1} T +  2 \lambda T_{j_1}   \Tr  ( R -1)   C 
 \nonumber\\
&-&  2 \lambda  \Tr_x [ R C_{j_1}  ] (x,x)  [  R C    ] (x,x)+ 2 \lambda \Tr [ ( R -1)C_{j_{1}}   ]T    -  \lambda\Tr [ ( R -1)C_{j_{1}}   ](x,x)[( R -1) C    ] (x,x)\biggr]
 \nonumber\\
&=& -3 \lambda  \int  e^{ V }  d\nu  (\sigma)  \Tr_x  [ (R -1) C_{j_{1}}   ] (x,x) [ ( R -1)C    ] (x,x) = \int  e^{ V }  d\nu  (\sigma)  A_{G( j_1)}(\sigma), \label{simpler}
\eea
where we recall that $\Tr $ means $\int d^2 x $, but we have also written more explicitly $\Tr_x $ for $\int d^2 x $ when we have more than one operator at coinciding $x$. 
This expression can be labeled by two order 1 resolvent graphs  corresponding to the two order 1 intermediate graphs $G_1$ and $G_2$ of Figure \ref{graphinter7}.
The $\sigma$ Wick contraction reconstructs a single former $\phi^4$ vertex, represented as a wavy line. The two c-propagators are  $(R-1)C_{j_1} $ and $(R-1)C$,
because they are both of the tadpole type. $C_{j_1}$ bears a red mark $j_1$, $C$ is a regular line without index, and we 
represent each $R-1$ as a dotted line. 

\begin{figure}[!t]
\begin{center}
{\includegraphics[width=10cm]{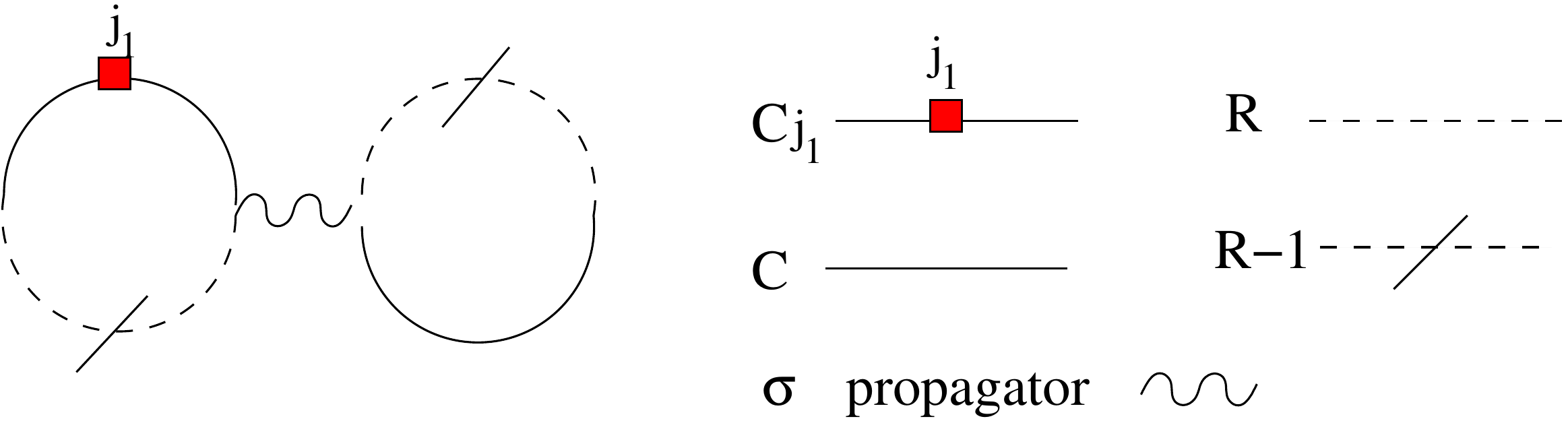}}
\end{center}
\caption{Minimal $G(j_1)$ resolvent graph: the single first order graph}
\label{graphinter1}
\end{figure}

The combinatoric weights of the two graphs are again 1 and 2.
Since they have again the same value (because of the flipping symmetry
or duality of the vertex) the result can be simplified as in \eqref{simpler} and expressed as a single
minimal resolvent graph decorated with a single mark $j_1$, hence corresponding to $\vert J \vert = 1$, which is
%The result is a sum over the two intermediate resolvent graphs,  decorated with
%a single mark $j_1$. Using the flipping symmetry we can reexpress the result as a single  
the graph $G( j_1)$ pictured in Figure \ref{graphinter1}. 
Remark that the factor $-3 \lambda$ can be understood as $-3 \times 2 (\lambda/2)$, where $\lambda/2$ is the coupling constant, (see \eqref{model})
3 is the ordinary combinatoric factor for the (single) order 1 vacuum $\phi^4$ graph, 
and the factor 2 comes from the two places in that graph where $C_{j_1}$ can be substituted to $C$.

Applying a second derivative term $\frac{d}{dt_{j_2}}$ we obtain
\bea  I_{12}&=& \int \frac{d}{dt_{j_1}}  \frac{d}{dt_{j_2}}   e^{ V} d\nu  (\sigma) =  -3 \lambda \int e^{ V } d\nu  (\sigma) \biggl(
 \Tr_x \bigl\{
 [(R -1) C_{j_{1}}   ] (x,x) [ ( R -1) C_{j_{2}}   ](x,x)  \nonumber  \\ 
 &-&2 i \sqrt \lambda \bigl(  [  R C_{j_{2}}  \sigma R  C_{j_{1}}  ]   (x,x)[  ( R -1) C  ] (x,x) + [( R -1) C_{j_{1}}   ] (x,x) [R C_{j_{2}}  \sigma R  C   ] (x,x)  
\bigr)
\bigr\}
\nonumber\\
&+& \Tr _{x,y}   [(R -1) C_{j_{1}}  ]  (x,x) [  (R -1)C  ](x,x)  \bigl[ 6 \lambda T_{j_2} T + 2i \sqrt{\lambda} T_{j_2} \sigma 
- i \sqrt{\lambda}   C_{j_{2}} \sigma ( R -1)  \bigr] (y,y)   \biggr) \label{derivok}
\eea
where $\Tr_{x,y} $  means $\int d^2 x d^2 y$ (beware to use the correct order for composition of operators!).

\begin{figure}[!t]
\begin{center}
{\includegraphics[width=12cm]{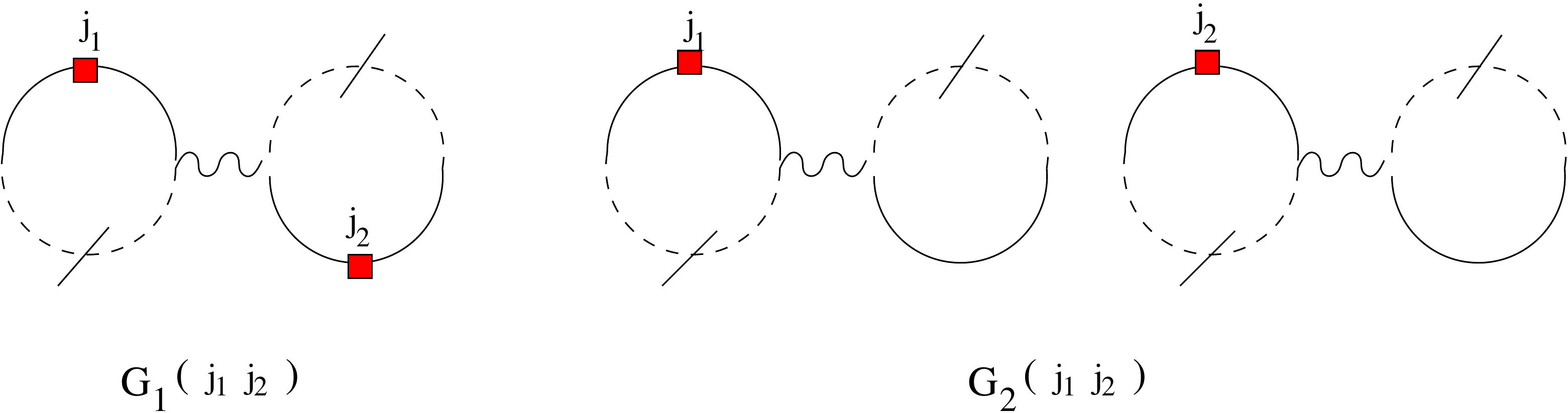}}
\end{center}
\caption{Minimal $\{j_1,j_2\}$ resolvent graphs: the first order graph and the second order disconnected graph}
\label{graphinter2}
\end{figure}

We can keep the term 
\bee 
 J_{12} =  - 3 \lambda  \int  e^{ V }  d\nu  (\sigma)   [ ( R -1) C_{j_{1}}   ]  (x,x) [  ( R -1) C_{j_{2}}  ] (x,x) =  \int  e^{ V }  d\nu  (\sigma)    A_{G_1( j_1,j_2)}(\sigma)
\ee
without further change. It is the only minimal $(j_1,j_2)$-decorated resolvent graph at order 1, pictured as $G_1(j_1,j_2)$ in Figure \ref{graphinter2}.

The contraction of the explicit $\sigma$ field in the other terms in \eqref{derivok} again allows 
for the cancellation of tadpoles, some obvious and others requiring attention.
To organize the computation we remark that in the last term, the contraction of the sigma field inside the same trace or to the exponential reproduces
the previous computation. Hence we obtain a relatively simple term which correspond to a disconnected
graph, namely the graph $G_2 ( j_1,j_2) $ in Figure \ref{graphinter2} whose value is
\bea K_{12}&=&9 \lambda^2 \int  e^{ V }  d\nu  (\sigma)   \Tr_x  [ (R -1) C_{j_{1}}   ] (x,x) [ ( R -1) C    ] (x,x)  \Tr_y  [( R -1) C_{j_{2}}   ] (y,y) [  ( R -1)C   ] (y,y) \nonumber\\
&=& \int  e^{ V }  d\nu  (\sigma) A_{G( j_1)}(\sigma)A_{G(j_2)}(\sigma).
\eea 
We remark that the integrand for a disconnected such resolvent graph factorizes over its connected components, as for usual Feynman graphs.

Then writing $I_{12} = J_{12} + K_{12} + L_{12}$ we obtain a lengthy sum of terms for $L_{12}$, which correspond to second order connected graphs (omitting the
trivial $\Tr_x$ and $\Tr_y$ integrations for simplicity):
\bea  L_{12}&=&   6 \lambda^2 \int e^{ V } d\nu  (\sigma) 
%\biggl(
\bigl\{ -2 T\bigl( [ R C_{j_{2}}  R  C_{j_{1}}    ](x,x) [(R-1) C] (x,x) +  [(R-1) C_{j_{1}} ] (x,x)  [ RC_{j_{2}}R C ] (x,x)  \bigr)
\nonumber\\
 &+&  [ ( R -1) C] (x,x) \bigl( [ RC ] (x,y) [RC_{j_{2}}   ] (y,x)  [( R -1) C_{j_{1}}  ] (y,y)
+ [R  C_{j_{1}}  ](x,y)  [R C_{j_{2}}  ]   (y,x) [  ( R -1)C   ] (y,y) \bigr)
\nonumber\\
 &+& 2 [ ( R -1)C_{j_{1}}  ] (x,x) \bigl( [RC  ] (x,y) [ R C] (y,y)[R C_{j_{2}}  ] (y,x) 
+ [ R  C ](x,y)  [R C_{j_{2}}  ]   (y,y) [  R C ] (y,x) \bigr)
\nonumber\\
 &+&2  [ (R -1)C  ] (x,x) \bigl( [R C_{j_{1}}  ] (x,y) [R C ] (y,y)  [R C_{j_{2}}  ] (y,x)
+ [ R C_{j_{1}}  ](x,y)  [R C_{j_{2}}  ]   (y,y) [ RC    ] (y,x) \bigr)
\nonumber\\
 &+&2  [R C ] (x,y) [RC_{j_{1}}   ] (y,x)   [R C ] (x,y) [RC_{j_{2}}   ] (y,x) +2  [RC  ] (x,y) [RC  ] (y,x)   [R C_{j_{1}} ] (x,y) [RC_{j_{2}}   ] (y,x) 
 \nonumber\\
 &-& 2   T_{j_2} \bigl(  [RC_{j_{1}} RC](x,x) [( R -1) C ] (x,x)    +   [ RCRC](x,x) [( R -1)  C_{j_{1}} ] (x,x)     \bigr)
\nonumber\\
&+&   [ ( R -1) C_{j_2} ] (x,x) \bigl(  [ R C ] (x,y) [ R C_{j_{1}}  ] (y,x)  [( R -1)  C ] (y,y) +
 [R C  ] (x,y)  [RC  ] (y,x)[( R -1)C_{j_1}  ] (y,y) \bigr)   \bigr\} .
 %\biggr)  
 \nonumber\\
 \label{deriv3}
 %\nonumber
\eea
We can now cancel the first and sixth lines of this equation (which contains tadpoles $T$ and $T_{j_2}$) respectively with the first term
of the resolvent expansion of the terms $C R (x,x)$ and 
$C_{j_2 R} (x,x)$ in lines 3 and 4. It renormalized the corresponding tadpoles, 
exactly transforming the corresponding $R$ factors into $(R-1)$ factors. Hence we get a tadpole-free expression:
\bea  L_{12}&=&   6 \lambda^2 \int e^{ V } d\nu  (\sigma) 
%\biggl(
\bigl\{ 
\nonumber\\
 &+&  [( R -1) C ] (x,x) \bigl( [  RC ] (x,y) [ R C_{j_{2}}  ] (y,x)  [( R -1)C_{j_{1}}  ] (y,y)
+ [R  C_{j_{1}}  ](x,y)  [RC_{j_{2}}   ]   (y,x) [  ( R -1)C   ] (y,y) \bigr)
\nonumber\\
 &+& 2 [( R -1)C_{j_{1}}   ] (x,x) \bigl( [RC  ] (x,y) [ (R-1) C] (y,y)[R  C_{j_{2}} ] (y,x) 
+ [  R C ](x,y)  [(R-1)C_{j_{2}}   ]   (y,y) [ RC   ] (y,x) \bigr)
\nonumber\\
 &+&2  [( R -1)C  ] (x,x) \bigl( [ R C_{j_{1}} ] (x,y) [(R-1) C ] (y,y)  [RC_{j_{2}}   ] (y,x)
+ [R  C_{j_{1}}  ](x,y)  [(R-1) C_{j_{2}}  ]   (y,y) [ R C   ] (y,x) \bigr)
\nonumber\\
 &+&2  [ RC ] (x,y) [R C_{j_{1}}  ] (y,x)   [RC  ] (x,y) [ R C_{j_{2}} ] (y,x) +2  [RC  ] (x,y) [R C ] (y,x)   [R C_{j_{1}} ] (x,y) [RC_{j_{2}}   ] (y,x) 
\nonumber\\
&+&   [( R -1)C_{j_2}  ] (x,x) \bigl(  [ R C ] (x,y) [R  C_{j_{1}}  ] (y,x)  [( R -1) C  ] (y,y) +
 [R C ] (x,y)  [RC  ] (y,x)[( R -1)C_{j_1}  ] (y,y) \bigr)    \bigr\} \nonumber\\
 \label{deriv4}
\eea
Using the flipping symmetry, we can simplify it slightly
and rewrite it as
\bea  L_{12}&=&   6 \lambda^2 \int e^{ V } d\nu  (\sigma)\bigl\{ 3 [( R -1)C  ] (x,x) [RC_{j_{1}}   ] (x,y) [ R C_{j_{2}} ] (y,x)  [ R -1)C( ] (y,y)
 \nonumber\\
 &+& 3 [( R -1)C  ] (x,x) [R C] (x,y) [RC_{j_{1}}   ] (y,x)  [( R -1)C_{j_{2}} ] (y,y) 
  \nonumber\\
 &+& 3[( R -1)C  ] (x,x) [RC ] (x,y) [RC_{j_{2}}   ] (y,x)  [( R -1) C_{j_{1}}] (y,y)
\nonumber\\
 &+&  3[( R -1)C_{j_{1}}   ] (x,x) [R C] (x,y) [R C ] (y,x)  [( R -1)C_{j_{2}} ] (y,y) 
\nonumber\\
 &+&2 [RC  ] (x,y) [R C_{j_{1}}  ] (y,x)   [RC  ] (x,y) [RC_{j_{2}}   ] (y,x) +2  [RC  ] (x,y) [R C ] (y,x)   [RC_{j_{1}}  ] (x,y) [RC_{j_{2}}   ] (y,x) 
  \bigr\} .
 \label{deriv5}
 \nonumber
\eea

\begin{figure}[!t]
\begin{center}
{\includegraphics[width=12cm]{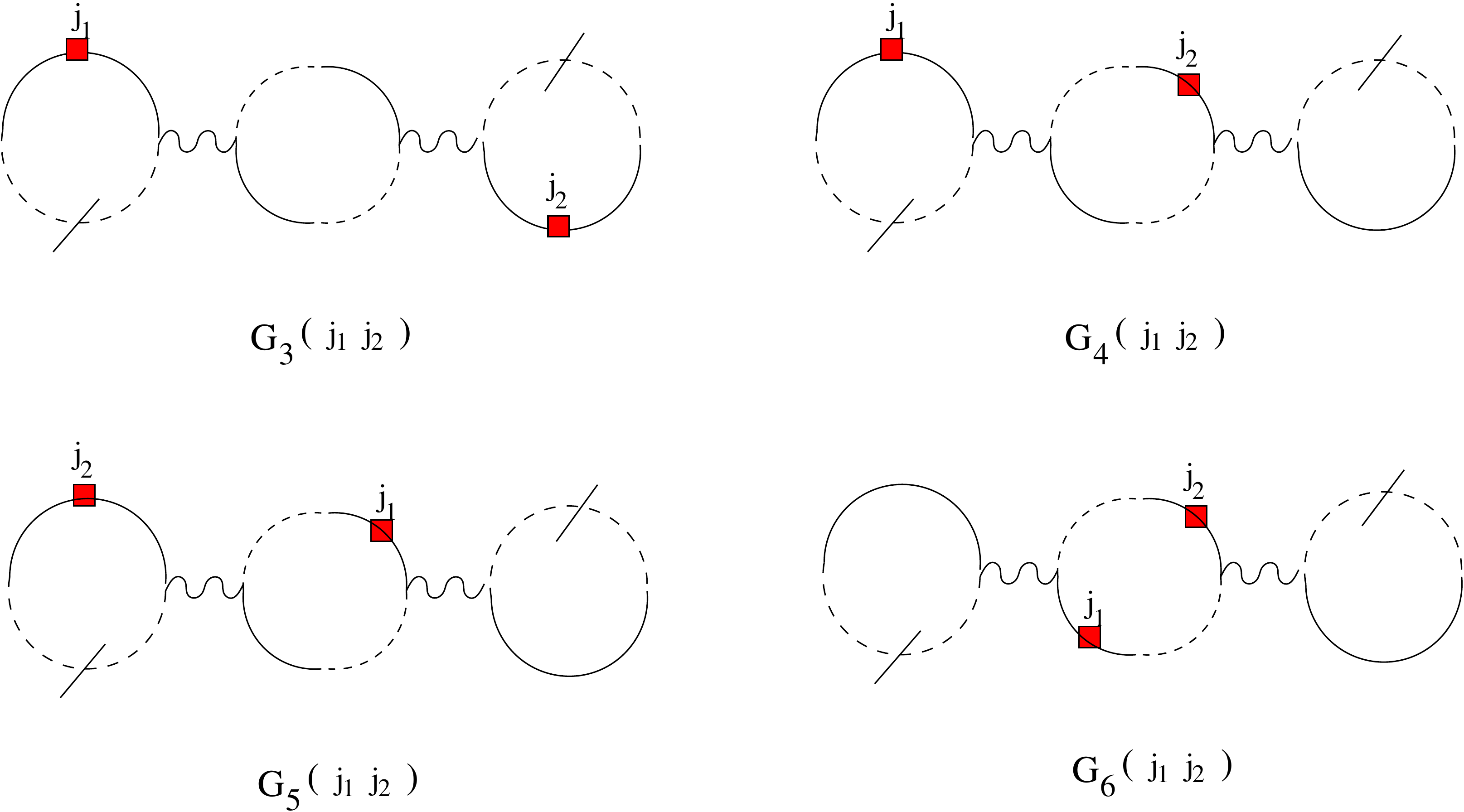}}
\end{center}
\caption{Minimal $(j_1,j_2)$-decorated resolvent graphs: second order connected graphs}
\label{graphinter3}
\end{figure}

\begin{figure}[!t]
\begin{center}
{\includegraphics[width=9cm]{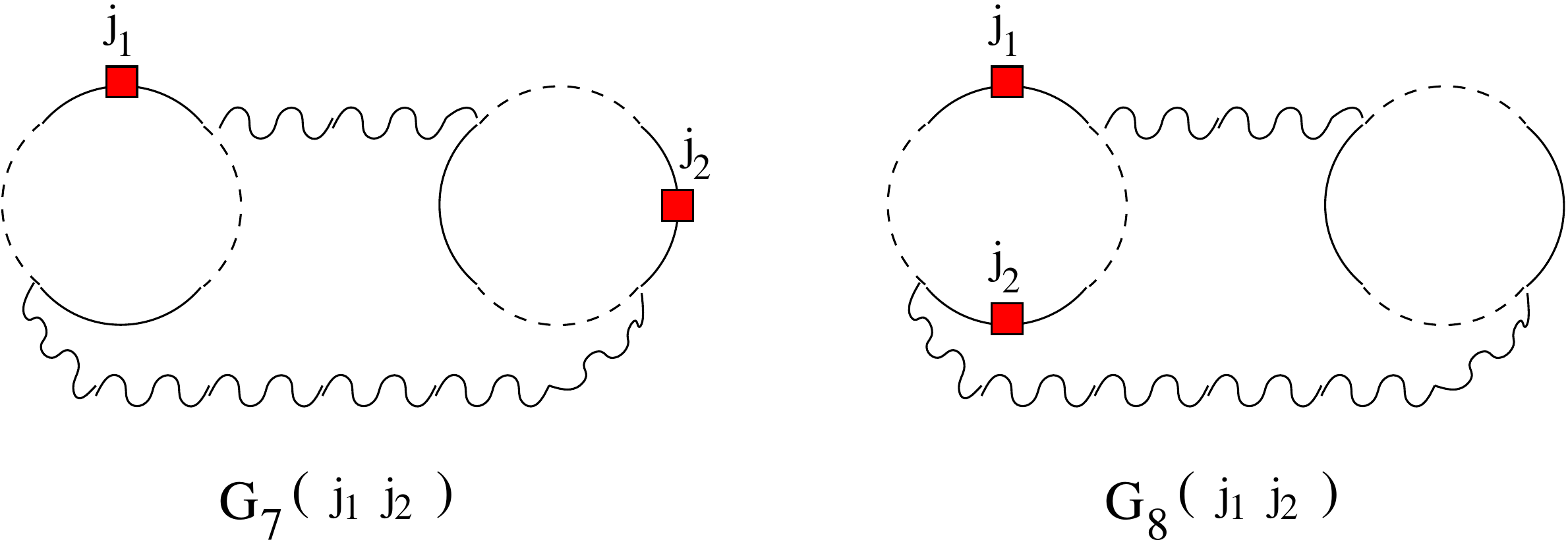}}
\end{center}
\caption{Minimal $(j_1,j_2)$-decorated resolvent graphs: second order connected graphs (end)}
\label{graphinter4}
\end{figure}

The result is  indexed by the graphs $G_i ( j_1,j_2) $, $i=3, \cdots 6$ in Figures \ref{graphinter2} and \ref{graphinter3} which have weight 18$\lambda^2$ 
and of graphs $G_i ( j_1,j_2) $, $i=7, 8$ in Figure \ref{graphinter4}, with weight
$12\lambda^2$. Remark that although there are a 
priori more intermediate field graphs at order 2 than shown in these figures, because of duality,
using the flipping symmetry of the $\sigma$ propagator we have reexpressed the result in terms of the fewer
graphs of Figures \ref{graphinter2}, \ref{graphinter3} and \ref{graphinter4}.

Collecting these results we have obtained the sum over all vacuum Feynman graphs of order 1 and 2 
with ordinary propagators $CR$ and tadpole propagators $C(R-1)$, in which one propagator is marked $j_1$ and the other marked $j_2$, in all possible
ways, with their natural symmetry factors, with just one further restriction, namely that every 
connected component of $G$ must bear at least one mark (see Figures \ref{graphinter2}-\ref{graphinter4}). 

The key step is the cancellation of the $\sigma $ fields associated to $T$ or $T_j$ factors with the tadpoles created 
by contractions of the $\sigma$ fields associated to $C_j$ or $C$ propagators to their own neighbor in the same loop.
This cannot be an accident. The intermediate field representation represents exactly the 
same graphs as the initial theory,  in which no tadpole can exist, and since the slice expansion tests blindly the presence
at first order of a given scale $j$ anywhere in the theory, either within graphs or counterterms, without \emph{any constraint}, this
cancellation \emph{must happen}. The next lemma states this for arbitrarily large $\vert J \vert$.

\begin{lemma} \label{testinglemma}
The general term of the testing expansion is 
\bee  Z^{j_{max}}(\lambda)=   \sum_{J \subset \cS}
\sum_{G \subset \cG (J)}  c_G
\int d\nu  (\sigma)     \prod_{j \in J (G)}  
\int_0^1 dt_j  \; \bigl[ e^{V (t) }  A^R_G (t, \sigma)   \bigr]_{t_j = 0 \; {\rm for} \; j \; \not\in J (G) } \label{testing1}
\ee 
where the sum over $\cG (J)$ runs over a set of minimal (vacuum) resolvent graphs (connected or not). 
The renormalized amplitudes $A^R_G$
are defined by \eqref{ampresren}, where the index $j(\ell)$ specifies the markings, that is restricts the c-propagator $\ell$ to 
be $C_j$ if that propagator bears the mark $j$. All marked propagators which belong to a short cycle (i.e. potential tadpole)
are \emph{renormalized}, hence accompanied by an $R-1$ resolvent factor, and not an $R$ factor. 
The c-propagators which do not bear any mark are equal to $C(t)$. 
\end{lemma}
\prf

The key step is to generalize the cancellation from \eqref{deriv3} to \eqref{deriv4}. By induction 
on $n=\vert J \vert $ we suppose Lemma \ref{testinglemma} holds up to $n$ slices $J=\{j_1, ... , j_n\}$, and we add
a new $d/dt_{j_{n+1}}$ and contract the explicit $\sigma$ fields it created. 
If this process creates a new disconnected graph, it is the first step computation \eqref{simpler} and we get
a new marked and renormalized connected component $G(j_{n+1})$ The other terms decompose into: 

A) terms with exactly one explicit $T_{j_{n+1}} $ counterterm, in which the $\sigma$ associated to $T_{j_{n+1}}$ has contracted into a resolvent of a
graph created by the previous steps,

B) terms with exactly one explicit $\Tr  RC_{j_{n+1}}$ ("potential tadpole" of scale $j+1$).

- other terms with no $T_{j_{n+1}}$ counterterm and no potential tadpole of any scale $j_1,   \cdots, j_{n+1}$.

Developing the potential tadpoles as $\Tr  RC_{j_{n+1}} = T_{j_{n+1}} + \Tr  (R-1)C_{j_{n+1}}$ 
transforms them into singular $T_{j_{n+1}}$ parts plus renormalized tadpoles $\Tr  (R-1)C_{j_{n+1}}$. The Lemma 
states that the $A$ terms exactly cancel the singular $B$ terms. This can be checked inductively by a painful analysis of cases
where $\sigma$ contract left or right. But we prefer a more abstract argument. Remark that 
we could apply our expansion to \emph{any} scale decomposition of the propagator into any finite subset of scales (not
necessarily \emph{contiguous}) and the algebra would always be the same. If we then put all $t_{j_k}$, $k=1, \cdots , n+1$, 
parameters to zero in every term after our $j_{n+1}$
expansion step and keep a last remaining (infrared) $C=C_0$ scale, what we obtain must be the $:\phi^4_2:$ theory with 
exactly a single propagator or tadpole
of each scale $j_1, \cdots, j_{n+1}$ and an arbitrary number of $C_0$ propagators and tadpoles. Keeping the scales $j_1, \cdots, j_n$
all fixed and letting $ j_{n+1} \to \infty$ 
proves that the $A$ and singular $B$ coefficients exactly cancel, otherwise the corresponding $:\phi^4_2:$ perturbative theory
restricted to that sequence of scales and to first order in $C_{j_1}, \cdots , C_{j_{n+1}}$ 
would not be perturbatively finite (as it must be).
\qed

With a little more work we could also prove that the terms with exactly one explicit $T$ counterterm 
and the singular part of the  terms  with exactly one explicit $\Tr  RC$ factor also exactly cancel each other, as seen in the computations above with one or two scales. 
Since we do not use this refinement in what follows, we leave such details to the reader.
Using the flipping symmetry we can also simplify the result to express it in terms of a minimal number of intermediate field graphs. 
This cannot play an important constructive role since the orbits at order $n$ have at most $3^n$ terms, 
hence do not change any of the constructive bounds. The combinatoric factors $c_G$ for the sum \eqref{testing1} are the natural ones so that putting resolvent factors
to 1 one recovers the right factors of $\phi^4$ perturbation theory up to order $\vert J\vert$, but the precise
rules to compute them can be a bit obscured by the use of the flipping symmetry and are not 
essential for constructive purpose (they can be bounded easily). Hence we leave them also to the reader.

\subsection{Factorization of Interaction}

Consider a fixed minimal  resolvent graph $G \in \cG$ corresponding to a non-empty set\footnote{Recall that an empty set of marks correspond
to the term 1 in the expansion of  $Z^{j_{max}}(\lambda)$.} of marks $J$ and to parameters $t_j$, $j \in J$. We need now also 
in \eqref{testing1} to factorize the action $e^{V(t)}$ into pieces attributed to each slice\footnote{If we were to interchange this factorization and the testing expansion, the cancellation of tadpoles with counterterms would not be exact.}.

In order to perform this factorization we shall attribute to each loop vertex the index of its highest c-propagator. To write explicitly the result, hence
the piece of the loop vertex attributed to slice $j$, we introduce an additional auxiliary parameter $u_j $, $j \in J$ which multiplies $t_j$ (we could also say that we substitute
$t_j (u_j) = u_j t_j$ to $t_j$) and we rewrite the interaction with cutoff $j$ as
\bee
V_{\le j} (t,\sigma) \equiv \Tr\biggl[3\lambda  T^2_{\le j}(t) + 2i \sqrt{\lambda} T_{\le j}(t) \sigma
-\frac{1}{2} \log_2[1+2i \sqrt{\lambda}C_{\le j}(t) \sigma ] \biggr]_{u_j = 1} .
\ee
The specific part of the interaction which should be attributed to the scale $j$ is the sum over all loop vertices with
at least one c-propagator at scale $j$ and all others at scales $\le j$. Hence it is  
\bea  V_{j} &\equiv&  V_{\le j}  - V_{\le j-1} =  V_{\le j} \vert_{u_j = 1} - V_{\le j} \vert_{u_j = 0}
\nonumber\\ &=& \int_0^1  du_j \frac{d}{du_j} \Tr   \biggl[ 3\lambda  T^2_{\le j}(t_j)  + 2i \sqrt{\lambda} T_{\le j}(t_j) \sigma 
-\frac{1}{2}    \log_2[1+2i \sqrt{\lambda}C_{\le j}(t_j) \sigma ] \biggr] \nonumber \\
&=&  \int_0^1 t_j du_j \Tr  \biggl[ 6 \lambda T_{j} T_{\le j} (t )  + 2i \sqrt{\lambda} T_{j} \sigma 
 - i \sqrt{\lambda}  [ R_{\le j}(t) -1 ]  C_{j} \sigma \biggr]  
\label{slicedinter}
\eea
where we recall that the resolvent is
\bee R_{\le j}(t) = ( 1+2i \sqrt{\lambda}C_{\le j}(t) \sigma)^{-1}.
\ee
Remark that since $t_j =0$ for $j \not\in J$, the total interaction can be written as 
\bee V_{\le j_{max}}  = \sum_{j =0}^{j_{\max}} V_j   =  \sum_{j \in J} V_j ,
\ee
since all terms $V_j$ with $j \not\in J$ are put to zero by the condition $t_j = 0$.

To each graph $G \in \cG(J)$ and $J\subset \cS$ is associated a partition into the (non-empty) connected components $G_1, \cdots , G_n$ of $G$ and an associated
partition $J_1, \cdots , J_n$ where $J_a \subset \cS$ is the (non-empty) set of marks present in $G_a$. We have $J_a \cap J_b= \emptyset$ for any
$a \ne b$. Since combinatoric weights of Feynman graph factor over connected components,
$c_G = \prod_a c_{G_a}$ and we can rewrite the result of the slice-testing expansion as

\bee  Z^{j_{max}}(\lambda)=   \sum_{n=0}^{\infty}  \frac{1}{n!}   \int d\nu  (\sigma)   \sum_{J_1, \cdots , J_n , \; J_a \cap J_b= \emptyset } 
\bigl[ \prod_{a =1}^n \prod_{j\in J_a}  \int_0^1 dt_j  \bigr] \prod_{a =1}^n \cI (J_a, t , \sigma) \label{testing21} 
\ee 
where 
\bee  \cI (J_a, t, \sigma) = \sum_{G_a \in \cC\cG (J_a) } c_{G_a}\bigl(  \prod_{j\in J_a} [  e^{V_j(t,\sigma) }]
A^R_{G_a}(t,\sigma) \bigr) \label{testing22} ,
\ee
$\cC\cG (J)$ being the set of \emph{connected} graphs in $\cG (J)$.
The $1/n!$ in \eqref{testing21} comes from summing over
\emph{ordered} sequences $J_1, \cdots ,  J_n$ in \eqref{testing21}. The sum over $n$ is in fact not infinite since the hardcore constraint
forces all terms to be zero for $n > 1 +j_{max}$. The term $n=0$ corresponds to the factor 1 in the sum (the normalization of the free theory).

We would like to factorize the sums over the blocks $J_a$. The functions $\cI$ are still coupled through
the $\sigma$ functional integral, the hardcore constraint $J_a \cap J_b= \emptyset $
and the $t$ parameters. Indeed the parameters $t$ in $\cI  (J, t, \sigma) $ are not restricted to those of the block
$J$ \footnote{We are very grateful to an anonymous referee for pointing this subtlety to us,
as we missed it in an earlier version of the paper.}. The first two issues are rather standard and in the next section we shall treat them
with the strategy of \cite{MLVE}. We now address the last issue through an \emph{auxiliary expansion}. 

Consider an integral such as 
\bee F =  \int  d\vec t_1 \cdots d \vec t_n  \prod_{a=1}^n f_a (\vec t_1,  \cdots \vec t_n)
\ee
We can introduce for each pair $1\le a <b \le n$ a pair interpolation parameters $x_{a,b}$ to 
simultaneously multiply the $t_b$ dependence of $f_a$ and the $t_a$
dependence of $f_b$, so we write 
\bee F =  \int d\vec t_1 \cdots d\vec t_n  \prod_{a=1}^n f_a (x_{1,a}\vec t_1,  \cdots, x_{a-1,a} \vec t_{a-1}, \vec t_a, x_{a,a+1} \vec t_{a+1},  \cdots  x_{a,n}\vec t_n) \vert_{x_{a,b} =1 \forall a,b}
\ee
The forest formula of \cite{BK,AR1} applied to $F$ gives
\bee F = \sum_\cF \int d\vec t_1 \cdots d\vec t_n  \int  dw_\cF  
\partial^\cF  \prod_{a=1}^n f_a (x_{1,i}\vec t_1,  \cdots, x_{a-1,a} \vec t_{a-1}, \vec t_a, x_{a,a+1} \vec t_{a+1},  \cdots  x_{a,n}\vec t_n) \vert_{x_{a,b} =x^\cF_{a,b} (\{w\}) \forall a,b}
\ee
where the sum is over all \emph{forests} $\cF$ on $(1, \cdots,n)$, $ \int dw_\cF $ means $\prod_{\ell =(a,b) \in \cF}  \int_0^1 dw_{a,b}$,
$ \partial^\cF $ means $ \prod_{\ell =(a,b) \in \cF} \frac{\partial}{\partial x_{a,b}}$  and $x^\cF_{a,b} \{w\}$ is as usual the infimum of the $w_{a',b'}$'s on the path 
of links $(a',b')$in $\cF$ joining $a$ to $b$ (0 if no such path exists) \cite{BK,AR1}.

Let us apply this formula to \eqref{testing21} and to $f_a =  \cI (J_a, t, \sigma) $, with $\vec t_a$ being the list of $\vert J_a \vert $ parameters
$t_j $ for $j \in J_a$.  We have to compute the action of $ \partial^\cF$ on $f$. Any derivative $\frac{\partial}{\partial x_{a,b}}$ in $\partial^\cF$
can act either on an $e^{V_j(t,\sigma) }$ factor or on the propagators 
resolvents of $A^R_{G_a}(t,\sigma)$. 
In the first case a derivative creates either a new loop vertex 
$\frac{\partial}{\partial t_j} V_{j'}  =  $  of any scale $j \in J_a$ for any  $j' \in J_b$ or the converse (exchange $a$ with $b$ in the sentence). 
each derived (crossed) $C$ propagator comes equipped with a $\sigma R$ factor.  
They can also act on the propagators $C_j $ for any  $j \in J_b$ in the propagators or resolvents of $A^R_{G_a}(t,\sigma)$, or the converse 
(exchange $a$ with $b$ in the sentence). Hence they always add a new propagator $C_j$ or $C_i$ or a tadpole $3 \lambda T_j T_{i}$
by hitting $3 \lambda T^2$. These new propagators or tadpoles are crossed to distinguish them form the previous marked ones.  
Note, although this is not essential, that the auxiliary expansion cannot act on the $2i \sqrt \lambda T_{j}$ term in \eqref{slicedinter},
since it contains a single frequency $j$, hence cannot be interpolated by any $x_{ab}$ parameter.

The important point is that for this auxiliary expansion we make absolutely no attempt to cancel any tadpoles with counterterms,
Such cancellations are no longer necessary because we already gathered extremely strong convergence factors for ultraviolet convergence 
through the marked renormalized propagators of the previous $t$ expansion, that can easily absorb a rather large number of logarithmic divergencies.

The result is a sum over new  larger blocks $J$ of indices, which are finite unions of $p$ previous blocks  $J_{a}$, 
those corresponding to the connected components of $\cF$, hence to 
trees $\cT$ of $\cF$ with $p-1$ edges. The associated modified amplitudes $ \cI\cM (J, t, \sigma)$ contain

\begin{itemize}

\item an integral $\prod_{j\in J}  \int_0^1 dt_j $ which we note $\int dt_J$ for short,

\item a sum over all partitions $\Pi$ of $J$ into finite disjoint non empty blocks $J = J_{a_1} \cup \cdots \cup J_{a_p} $,

\item a sum over all trees $\cT$ on $\{1,  p\}$,

\item an integral $\int dw_\cT  $ over $w$ parameters, 

\item the exponential of the interaction factor $ \prod_{j\in J} [  e^{V_j(t,\sigma) }] $, where the $t$'s
are now multiplied by their appropriate $x^\cT_{a,b} (\{w\})$ factors,

\item a sum over $G$ in the set $ \cC\cG\cM (J, \Pi, \cT)$ indexing all the ways the derivatives in $\partial^\cT$ acted,
of a corresponding modified amplitude $A^{R,M}_{G}(t,\sigma)$. The modified graph $G$ can be pictured as
containing marked renormalized propagators $C_j$, crossed propagators and/or counterterms plus ordinary propagators and $R$ or $R-1$ resolvents. 
Its amplitude is obtained by collecting all the terms brought down from the exponential both by the slice testing expansion which created the blocks $J_{a_1}, \cdots , J_{a_p} $
and by the auxiliary expansion which joined $J_{a_1}, \cdots , J_{a_p} $. The corresponding modified graphs
have the following characteristics

\begin{itemize}

\item they are no longer necessarily connected,

\item they contain exactly one renormalized marked propagator $C_j$ for every $j \in J$ (hence equipped with an $R-1$ factor if in a short 
loop of length 1)

\item they contain at most $\vert J \vert -1$ crossed propagators or $T_j$ counterterms, each with a scale index $j \in J$. A 
crossed propagator is preceded by an $R$ or $R-1$ resolvent. 

\item their order of perturbation (number of explicit $\lambda$ factors not inside resolvents) is bounded by $4\vert J \vert $

\end{itemize}

\end{itemize}

\begin{figure}[!t]
\begin{center}
{\includegraphics[width=6cm]{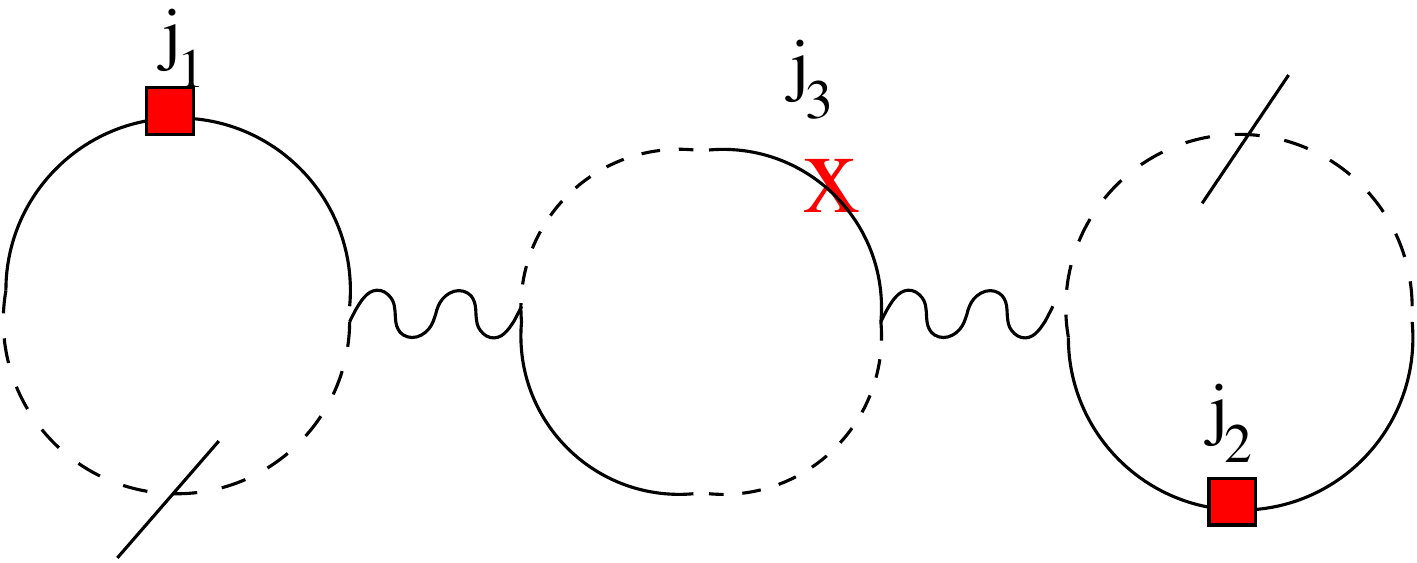}} \hskip1cm {\includegraphics[width=4cm]{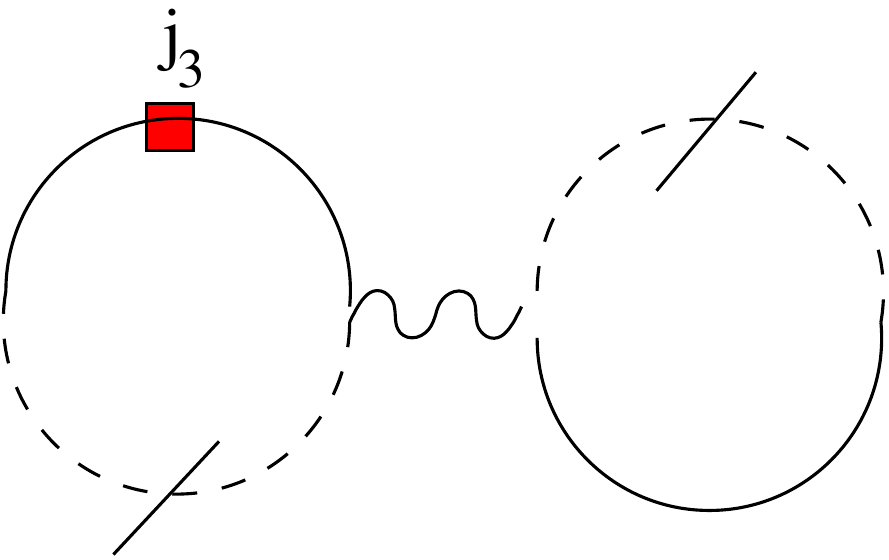}}
\end{center}
\caption{A modified graph connecting the block $J_a = \{ j_1, j_2\}$ to the block $J_b = \{j_3\}$}
\label{graphmodified}
\end{figure}

This is illustrated in Figure \ref{graphmodified}, in which two blocks $J_a = \{ j_1, j_2\}$ and $J_b = \{j_3\}$ are 
linked by a $d/dt_{j_3}$, acting in fact on a graph $G_3(j_1, j_2)$ of Figure \ref{graphinter3}. It decorated that graph with a crossed propagator of scale $j_3$.
The resulting graph for the new block $\{ j_1, j_2, j_3\}$ is disconnected and still bears three marked renormalized propagators, one
for each scale of $\{ j_1, j_2, j_3\}$. This is enough to ensure excellent ultraviolet
convergence of the new block, even if we do not care (during auxiliary expansion steps) to compensate short cycles of length 1 (potential tadpoles) 
with their associated crossed counterterms.

We can therefore write a modified formula similar to \eqref{testing22}, with hopefully clear notations
\bee  Z^{j_{max}}(\lambda)=   \sum_{n=0}^{\infty}  \frac{1}{n!}   \int d\nu  (\sigma)   \sum_{J_1, \cdots , J_n , \; J_a \cap J_b= \emptyset } 
\prod_{a =1}^n   \cI\cM (J_a, \sigma) \label{testing23} 
\ee 
where 
\bee  \cI\cM (J,  \sigma) = \sum_{\Pi}  \sum_{\cT}  \int dt_J  \int dw_\cT
\sum_{G \in \cC\cG\cM (J, \Pi , \cT) }    c_{G} \bigl(  \prod_{j\in J} [  e^{V_j(t,w,\sigma) }]
A^{R,M}_{G}(t,w,\sigma) \bigr) \label{testing24} .
\ee
The $w$ dependence means that all $t$ factors are multiplied by their appropriate unessential $x^\cT_{a,b} (\{w\})$ factors.

In \eqref{testing24}Êthe $t$ integrals now  have obviously factorized over the new blocks. That was the sole purpose of the auxiliary expansion.
We can now encode the hardcore constraints $ J_a \cap J_b= \emptyset $ through Grassmann numbers as
\bea  Z^{j_{max}}(\lambda) &=&  \sum_{n=0}^{\infty}  \frac{1}{n!}   \int d\nu  (\sigma)  d \mu_{\mathbb{I}_{\cS}} ( \bar \chi ,\chi ) 
\prod_{a =1}^n \biggl(  \sum_{J_a \subset \cS}   \sum_{\Pi}  \sum_{\cT}  \int dt_{J_a}  \int dw_\cT  
\nonumber \\
&&\sum_{G \in \cC\cG\cM (J, \Pi , \cT)  } c_{G} \bigl(  \prod_{j\in J_a} [ - \bar \chi_j  e^{V_j(t,w,\sigma) }  \chi_j ]
A^{R,M}_{G}(t,w,\sigma) \biggr) 
 \label{testing3}
\eea
where $d \mu_{\mathbb{I}_{\cS}}(\bar \chi ,\chi ) = \prod_{j=0}^{j_{max}} d\bar \chi_j d\chi_j \; e^{-\bar \chi_j \chi_j}$ is the standard normalized Grassmann 
Gaussian measure with covariance $\mathbb{I}_{\cS}$, the $1+j_{max}$ by $1+j_{max}$  identity matrix. Indeed Grassmann Gaussian variables automatically 
implement the hardcore constraints, and saturate the Grassmann pairs for $j \not \in J= \cup_a J_a$.

We can remark that  \eqref{testing3} is the developed expansion for a new type of vertex $W$ which is a sum over slice-subsets $J$:
\bea  Z^{j_{max}}(\lambda) &=& \int d\nu  (\sigma)  d \mu_{\mathbb{I}_{\cS}}(\bar \chi ,\chi )   e^{ W } , \label{testing4}  \\
W &=& \sum_{J \subset \cS}  \sum_{\Pi}  \sum_{\cT} \int dt_J \int dw_\cT   \sum_{G \in \cC\cG\cM (J, \Pi , \cT)  } 
 c_{G} \biggl(  \prod_{j\in J} [ - \bar \chi_j  e^{V_j(t,w,\sigma) }  \chi_j ]
A^{R,M}_{G}(t,w,\sigma) \biggr)\label{testing5}
\eea
To distinguish $W$, which contains potentially infinitely many vertices of type $V$, we call it an \emph{exp-vertex}.

From now on and until subsection \ref{finalsubsec}, in order not to distract the reader we drop in the writing of $W$ all the auxiliary expansion
$ \sum_{\Pi}  \sum_{\cT}  $ and $w$ factors and the $M$ index which recall that graphs $G$ can contain crossed propagators and/or crossed $T$ counterterms.
Indeed these auxiliary factors play absolutely no role, until we return to them, at the end of subsection 
\ref{finalsubsec},Êto see that they only trivially modify 
the unessential constants of the final bounds.

This ends the preparation phase and we can now proceed to the MLVE expansion proper.

\section{The Multiscale Loop Vertex Expansion}
\label{multilve}

We perform now the two-level jungle expansion defined in \cite{MLVE}, starting from \eqref{testing3} which is the developed form of \eqref{testing4}.
For completeness we summarize the main steps, referring to \cite{MLVE} for details.

The first step introduces Bosonic replicas for all the exp-vertices in \eqref{testing3}. Noting $\cW = \{1, \cdots , n\}$ the set of labels for these 
exp-vertices, we have 
\bee
Z^{j_{max}}(\lambda)= \sum_{n=0}^\infty \frac{1}{n!}  \int d\nu_{\cW}(\sigma, \bar \chi, \chi) \;  \prod_{a=1}^n  W_a  (\sigma_a, \bar \chi, \chi ) \; ,
\ee
so that each vertex $W_a$ has now its own Bosonic field $ \sigma^a  $. The replicated measure is completely degenerate between replicas:
\bea
d\nu_{\cW}  &=& d\nu_{\bbone_\cW} (\{   \sigma_a\}) \;  d\mu_{\mathbb{I}_\cS  } (\bar \chi_j , \chi_j)   \\
W_a  (\sigma_a, \bar \chi, \chi )   &=&  
\sum_{J \subset \cS}  \int dt_J   \sum_{G \in \cC\cG(J)  } 
 c_{G} \biggl(  \prod_{j\in J} [ - \bar \chi_j  e^{V_j(t,\sigma_a) }  \chi_j ]
A^{R}_{G}(t,\sigma_a) \biggr).
\eea
where $\bbone_\cW$ is the $n$ by $n$ matrix with coefficients 1 everywhere.

The obstacle to factorize the functional integral $Z$ over vertices and to compute $\log Z$ lies in the Bosonic degenerate blocks $\bbone_\cW$ and 
in the Fermionic fields which couple the vertices $W_a$. In order to remove these two obstacles we need to apply {\it two} successive
forest formulas \cite{BK,AR1}, one Bosonic, the other Fermionic. 

To analyze the block $\bbone_\cW$ in the measure $d \nu$ 
we introduce coupling parameters $x_{ab}=x_{ba}, x_{aa}=1$ between the Bosonic vertex replicas
and obtain a new sum over forests. Representing Gaussian integrals as derivative operators
as in \cite{MLVE} we have
\bee
Z^{j_{max}}(\lambda) = \sum_{n=0}^\infty \frac{1}{n!} 
\Bigl[ e^{\frac{1}{2} \sum_{a,b=1}^n x_{ab} 
\frac{\partial}{\partial \sigma_a}\frac{\partial}{\partial \sigma_b} 
   +  \sum_{j=0}^{j_{\max}} \frac{\partial}{\partial \bar \chi_j  } \frac{\partial}{\partial \chi_j } } \; 
   \prod_{a=1}^n  W_a  (\sigma_a, \bar \chi, \chi )
   \Bigr]_{ \genfrac{}{}{0pt}{}{  \sigma, \chi,  \bar\chi  =0}{x_{ab}=1 }} \;.
\ee
The next step applies the standard Taylor forest formula of \cite{BK,AR1} to the $x$ parameters. 
We denote by $\cF_{B}$ a Bosonic forest with $n$ vertices labelled $\{1,\dots n\}$.
It means an acyclic set of edges over $\cW$. For
$\ell_{B}$ a generic edge of the forest we denote by $a(\ell_B), b(\ell_B)$ the end vertices of $\ell_B$. The result of the Taylor forest formula is:
\beann
&&  Z^{j_{max}}(\lambda) = \sum_{n=0}^\infty \frac{1}{n!} \sum_{\cF_{B}} \int_{0}^1 \Bigl( \prod_{\ell_B\in \cF_B } dw_{\ell_B} \Bigr)
\; \;  \Bigg[  e^{\frac{1}{2} \sum_{a,b=1}^n X_{ab}(w_{\ell_B})  
  \frac{\partial}{\partial \sigma_a}\frac{\partial}{\partial \sigma_b} 
   +  \sum_{j=0}^{j_{\max}} \frac{\partial}{\partial \bar \chi_j  } \frac{\partial}{\partial \chi_j  } }
\crcr
&& \qquad \qquad \qquad \times 
   \prod_{\ell_B \in \cF_{B}} \Bigl(  
   \frac{\partial}{\partial \sigma_{a(\ell_B)}}\frac{\partial}{\partial \sigma_{b(\ell_B)}}  \Bigr) \;
 \prod_{a=1}^n  W_a  (\sigma_a, \bar \chi, \chi )\Bigg]_{ \sigma ,\chi , \bar\chi =0 } \; ,
\eeann
where $X_{ab}(w_{\ell_B})$ is the infimum over the parameters $w_{\ell_B}$ in the unique path
in the forest $\cF_B$ connecting $a$ to $b$. This infimum is set to 
$1$ if $a=b$ and to zero if $a$ and $b$ are not connected by the forest \cite{BK,AR1}.

The forest $\cF_B$ partitions the set of vertices into blocks $\cB$ corresponding to its connected components. In each such block the 
edges of $\cF_B$ form a spanning tree. Remark that such blocks can be reduced to bare vertices. Any vertex $a$
belongs to a unique Bosonic block $\cB$.
Contracting every Bosonic block to an ``effective vertex'' we obtain a graph which we denote $\{n\}/\cF_B$.

The next step introduces replica Fermionic fields $\chi^{\cB}_j$ for these blocks of $\cF_B$ (i.e. for the effective vertices 
of $\{n\}/\cF_B$) and replica coupling parameters $y_{\cB\cB'}=y_{\cB'\cB}$. 
The last step applies (once again) the forest formula, 
this time for the $y$'s, leading to a set of Fermionic edges $\cL_F$ forming a forest 
in $\{n\}/\cF_B$ (hence connecting Bosonic blocks). Denoting $L_{F} $ a generic Fermionic edge connecting blocks and $\cB(L_F), \cB'(L_F) $ 
the end blocks of the Fermionic edge $L_F$ we
follow exactly the same steps than in \cite{MLVE} 
and obtain a two level-jungle formula \cite{AR1}. It writes
\bee
Z^{j_{max}}(\lambda)  =  \sum_{n=0}^\infty \frac{1}{n!}  \sum_{\cJ} \;
 \;  \int dw_\cJ  \;  \int d\nu_{ \cJ}  
\quad   \partial_\cJ   \Big[ \prod_{\cB} \prod_{a\in \cB}      W_{a}   (   \sigma_a , \chi^{ \cB } , \bar \chi^{\cB}  )
  \Big] \; ,
\ee
where
\begin{itemize}

\item the sum over $\cJ$ runs over all two-level jungles,
hence over all ordered pairs $\cJ = (\cF_B, \cF_F)$ of two (each possibly empty) 
disjoint forests on $\cW$, such that $\cF_B$ is a forest, $\cF_F$ is a forest and
$\bar \cJ = \cF_B \cup \cF_F $ is still a forest on $\cW$. The forests $\cF_B$ and $\cF_F$ are the Bosonic and Fermionic components of $\cJ$.
 
\item  $\int dw_\cJ$ means integration from 0 to 1 over parameters $w_\ell$, one for each edge $\ell \in \bar\cJ$, namely
$\int dw_\cJ  = \prod_{\ell\in \bar \cJ}  \int_0^1 dw_\ell  $.
There is no integration for the empty forest since by convention an empty product is 1. A generic integration point $w_\cJ$
is therefore made of $\vert \bar \cJ \vert$ parameters $w_\ell \in [0,1]$, one for each $\ell \in \bar \cJ$.

\item 
\bee \partial_\cJ  = \prod_{\genfrac{}{}{0pt}{}{\ell_B \in \cF_B}{\ell_B=(a,b)}} \Bigl(
\frac{\partial}{\partial \sigma_a}\frac{\partial}{\partial \sigma_b} \Bigr)
\prod_{\genfrac{}{}{0pt}{}{\ell_F \in \cF_F}{\ell_F=(d,e) } } \sum_{j_{\ell_F} =0}^{j_{max}} \Big(
   \frac{\partial}{\partial \bar \chi^{\cB(d)}_{j_{\ell_F} } }\frac{\partial}{\partial \chi^{\cB(e)}_{j_{\ell_F} } }+ 
    \frac{\partial}{\partial \bar \chi^{ \cB( e) }_{j_{\ell_F} } } \frac{\partial}{\partial \chi^{\cB(d) }_{j_{\ell_F}  } }
   \Big) \; ,
\ee
where $ \cB(d)$ denotes the Bosonic block to which the vertex $d$ belongs. 

\item The measure $d\nu_{\cJ}$ has covariance $ X (w_{\ell_B}) $ on Bosonic variables and $ Y (w_{\ell_F}) \otimes \mathbb{I}_\cS  $  
on Fermionic variables, hence
\bee
\int d\nu_{\cJ} F = \biggl[e^{\frac{1}{2} \sum_{a,b=1}^n X_{ab}(w_{\ell_B })  \frac{\partial}{\partial \sigma_a}\frac{\partial}{\partial \sigma_b} 
   +  \sum_{\cB,\cB'} Y_{\cB\cB'}(w_{\ell_F})  \sum_{j_{\cB, \cB'} \in \cS}
   \frac{\partial}{\partial \bar \chi_{j_{\cB, \cB'}}^{\cB} } \frac{\partial}{\partial \chi_{j_{\cB, \cB'}}^{\cB'} } }   F \biggr]_{\sigma = \bar\chi =\chi =0}\; .
\ee

\item  $X_{ab} (w_{\ell_B} )$  is the infimum of the $w_{\ell_B}$ parameters for all the Bosonic edges $\ell_B$
in the unique path $P^{\cF_B}_{a \to b}$ from $a$ to $b$ in $\cF_B$. This infimum is set to zero if such a path does not exists and 
to $1$ if $a=b$. 

\item  $Y_{\cB\cB'}(w_{\ell_F})$  is the infimum of the $w_{\ell_F}$ parameters for all the Fermionic
edges $\ell_F$ in any of the paths $P^{\cF_B \cup \cF_F}_{a\to b}$ from some vertex $a\in \cB$ to some vertex $b\in \cB'$. 
This infimum is set to $0$ if there are no such paths, and to $1$ if such paths exist but do not contain any Fermionic edges.

\end{itemize}

Remember that a main property of the forest formula is that the symmetric $n$ by $n$ matrix $X_{ab}(w_{\ell_B})$ 
is positive for any value of $w_\cJ$, hence the Gaussian measure $d\nu_{\cJ} $ is well-defined. 

Since the slice assignments, the fields, the measure and the integrand are now 
factorized over the connected components of $\bar \cJ$, the logarithm of $Z$ is easily computed as exactly the same sum but restricted 
to two-levels spanning trees:
\bea \label{treerep}  
\log Z^{j_{max}}(\lambda)=  \sum_{n=1}^\infty \frac{1}{n!}  \sum_{\cJ \;{\rm tree}} \; \;  \int dw_\cJ  \;  \int d\nu_{ \cJ}  
\quad   \partial_\cJ   \Big[ \prod_{\cB} \prod_{a\in \cB}   \Bigl(     W_{a}   (   \sigma_a , \chi^{ \cB } , \bar \chi^{\cB}  )\Bigr)    \Big] \; , 
\eea
where the sum is the same but conditioned on $\bar \cJ = \cF_B \cup \cF_F$ being a \emph{spanning tree} on $\cW= [1, \cdots , n]$.
The main result is the convergence of this representation uniformly in $j_{max}$ for $\lambda$ in a certain domain, 
allowing to perform in this domain the ultraviolet limit of the theory. More precisely

\begin{figure}[!t]
\begin{center}
{\includegraphics[width=0.2\textwidth]{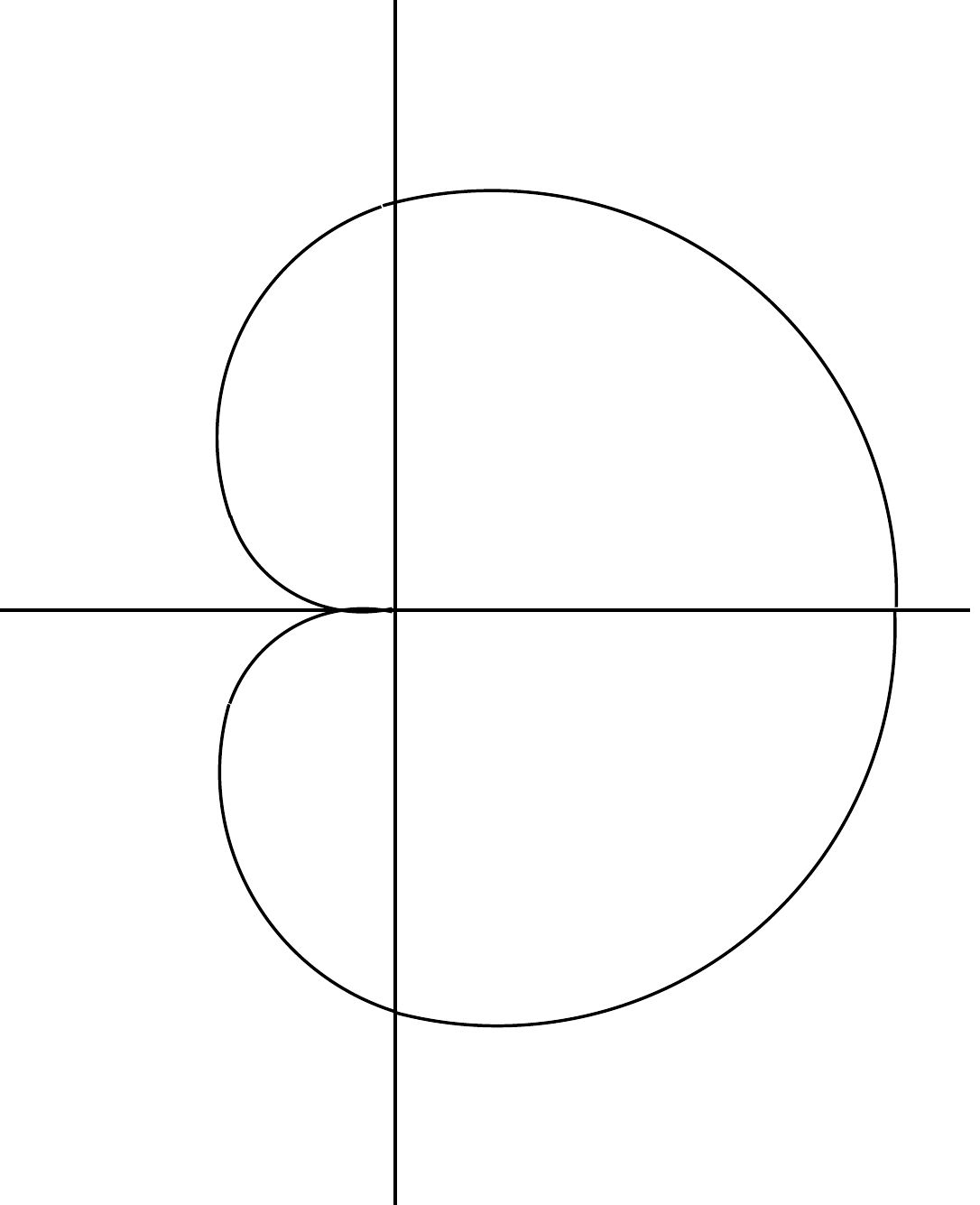}}
\end{center}
\caption{A Cardioid Domain}
\label{cardio}
\end{figure}

\begin{theorem} \label{thetheorem} Fix  $\rho >0$ small enough.
The series \eqref{treerep} is absolutely convergent, uniformly in $j_{max}$, for $\lambda$ in the small open cardioid domain $\cC ard_\rho$
defined by $\vert \lambda \vert < \rho \cos^2 [({\rm Arg} \; \lambda )/2]$ (see Figure \ref{cardio}).
Its ultraviolet limit $\log Z (\lambda) = \lim_{j_{max}  \to \infty}  \log Z^{j_{max}}(\lambda)$ is therefore well-defined and
analytic in that cardioid domain; furthermore it is the Borel sum
of its perturbative series in powers of $\lambda$. 
\end{theorem}
  
\section{The Bounds}

\subsection{Grassmann Integrals}

Each vertex $W_a$ is developed as a sum over $J_a$ and over connected graphs $G_a$ which contains the marks of $J_a$ exactly once.
It is then important to emphasize explicitly the constraints that the scale subsets $J_a$ obey to a \emph{hard core constraint inside each block}.
Indeed if there was any non empty such intersection one would integrate twice with respect to a same Grassmann variable corresponding to that block and to the same frequency.
Hence without changing the value of the expansion we can factor out a term $\prod_{\cB} \prod_{a,b\in \cB} \bbone( J_a \cap J_b =\emptyset) $,
where the function $\bbone(E)$ means the characteristic function of the event $E$.

The Grassmann Gaussian part of the functional integral \eqref{treerep} is then treated as in \cite{MLVE}, resulting in a similar computation. Let us 
for the moment fix the forest $\cF_F$, hence also the two ends $a,b$ of each Fermionic link $\ell_{\cF}$,
and let us also fix the scale $j_{\ell_F}$ off each Fermionic link $\ell_F$. All these data will be summed later. Definng as in \cite{MLVE} the  
natural $n$ by $n$ extension of the matrix $Y_{\cB, \cB'} $ by
${\bf Y}_{ab} = Y_{\cB(a) \cB(b)} (w_{\ell_F})$, 
 we can evaluate
\bea\label{eq:grassmaint}
\int \biggl[ \prod_{\cB} \prod_{j}  ( d  \bar \chi^{\cB}_{j}  d  \chi^{\cB}_{j}   ) \biggr]
 e^{  - \sum_{j=0}^{j_{max}}    \bar \chi^{\cB}_{j} Y_{\cB\cB'}(w_{\ell_F})   \chi^{\cB'}_{j}  }  
\prod_{\genfrac{}{}{0pt}{}{\ell_F \in \cF_F}{\ell_F=(a,b) } } 
\Big(    \chi^{\cB(a)}_{j_{\ell_F} } \bar  \chi^{\cB(b)}_{j_{\ell_F} }  + 
     \chi^{ \cB( b) }_{j_{\ell_F} }  \bar  \chi^{\cB(a) }_{j_{\ell_F} } \Big) \nonumber
 \\ =   
 \Bigl(\prod_{\cB} \prod_{a,b\in \cB} \bbone( J_a \cap J_b =\emptyset) \Bigr)
 \Bigl( {\bf Y }^{\hat b_1 \dots \hat b_k}_{\hat a_1 \dots \hat a_k}  + 
 {\bf Y }^{\hat a_1 \dots \hat b_k}_{\hat b_1 \dots \hat a_k}+\dots + {\bf Y }_{\hat b_1 \dots \hat b_k}^{\hat a_1 \dots \hat a_k}   \Bigr) \; ,
\eea 
where $k = \vert \cF \vert $ and the sum runs over the $2^k$ ways to exchange the ends $a_i$ and $b_i$ of each $\ell_F$,
and the $Y$ factors are (up to a sign) the minors of $Y$ with the lines $b_1\dots b_k$ and the columns $a_1\dots a_k$ deleted.
The most important factor in \eqref{eq:grassmaint} is  $\prod_{\cB} \prod_{a,b\in \cB} \bbone( J_a \cap J_b =\emptyset) $
which ensures the disjointness of the slices in each block.

Positivity of the $Y$ covariance means as usual that the $Y$ minors are all bounded by 1 \cite{AR2,MLVE}, namely
for any $a_1,\dots a_k$ and $b_1,\dots b_k$,
\bee
\Big{|}  {\bf Y }^{\hat a_1 \dots \hat b_k}_{\hat b_1 \dots \hat a_k} \Big{|}\le 1 \; .
\ee

\subsection{Bosonic Integrals}
\label{bosointegr}

The main problem is now the evaluation of the Bosonic integral $\int d\nu (\sigma)$ in \eqref{treerep}. Since it factorizes over the Bosonic blocks,
it is sufficient to bound separately this integral in each fixed block $\cB$. Consider such a block $\cB$,
and the fixed set of slice-subsets and graphs $S_a, G_a$ of that block. We shall define $J(\cB) = \cup_{a \in \cB}  J_a$,
hence it is the set of slices present in the block $\cB$ (remember the factor $ \prod_{a,b\in \cB} \bbone( J_a \cap J_b =\emptyset) $
which ensures that the $J_a$ are all disjoint for $a \in \cB$).
 
In the block, the Bosonic forest $\cF_B$ restricts to a Bosonic tree
$\cT_{\cB}$, 
and the Bosonic Gaussian measure  $d \nu $ restricts to $d \nu_\cB$ defined by
\bee
\int d \nu_\cB  F_\cB = \biggl[ e^{\frac{1}{2} \sum_{a,b\in \cB} X_{ab}(w_{\ell_B }) 
 \frac{\partial}{\partial \sigma_a}\frac{\partial}{\partial \sigma_b}} F_\cB \biggr]_{\sigma =0}.
\ee
The Bosonic integrand is obtained by evaluating the action of the coupling derivatives in
\bea \label{manyder}  F_\cB &=&  \prod_{a\in \cB}    \bigl[ \prod_{e \in E^a_\cB} \bigl( \frac{\partial}{\partial \sigma_e} \bigr)  W_{j_a}   \bigr]\nonumber\\
&=&  \prod_{a\in \cB}    \biggl[ \prod_{e \in E^a_\cB} \bigl( \frac{\partial}{\partial \sigma_e} \bigr)      \prod_{j\in J_a} \bigl[ \int_0^1 dt_j  e^{V_j(t,\sigma_a) }  \bigr] A_{G_a}(t,\sigma_a)   \biggr]\
\eea
where $E^a_\cB$ runs over the set of all edges in $\cT_{\cB}$ which end at vertex $a$, hence $\vert E^a_\cB \vert = d_a( \cT_{\cB} ) $, the
degree or coordination of the tree $\cT_{\cB}$ at vertex $a$. 
The  derivatives $\prod_{e \in E^a_\cB}$ act\footnote{
When the block $\cB$ is reduced to a single vertex $a$, there is no derivative to compute and the integrand reduces simply to 
$F_\cB  =   \prod_{j\in J_a} \bigl[ \int_0^1 dt_j  e^{V_j(t,\sigma_a) }  \bigr] A_{G_a}(t,\sigma_a)  $. This case is easy.} either on the amplitudes $A_{G_a}$ or derive new loop vertices 
from the exponential $\prod_{j \in J_a} e^{V_j (\sigma_a)}$.

When $\cB$ has more than one vertex, since $\cT_{\cB}$ is a tree, each vertex $a \in \cB$ is touched by at least one derivative. 
We can evaluate the derivatives in \eqref{manyder} through the Fa\`a di Bruno formula:
\bee
\bigl( \prod_{e\in E} \frac{\partial}{\partial \sigma_e}  \bigr)   f\bigl( g( \sigma ) \bigr) =   \sum_{\pi } f^{|\pi|}\bigl( g( \sigma ) \bigr) \prod_{B\in \pi}  \bigl[ \bigl( \prod_{e\in B} \frac{\partial}{\partial \sigma_e} \bigr)  g (\sigma)\bigr]  \; ,
\ee
where $\pi$ runs over the partitions of the set $S$ and $B$ runs through the blocks of the partition $\pi$. For the purpose of this paper we won't need to
evaluate too precisely the result, but let us remark that
 
\begin{itemize}
\item the exponential $\prod_{j \in J_a} e^{V_j (\sigma_a)}$ cannot disappear since the exponential function is its own derivative,

\item the derivatives which act on the graph integrand $A_{G_a}(t,\sigma_a) $ must act on the $R$ resolvent factors and create therefore new propagators sandwiched by resolvents,
through $\frac{\partial}{\partial \sigma} R =  2i \sqrt \lambda R C R$,

\item the derivatives which act on the exponential create new loop vertices of the type $ i \sqrt \lambda C (R-1) $ or $T_j$ counterterms.
\end{itemize}

We do not try to compensate eventual tadpoles $T_j$ created by these $\frac{\partial}{\partial \sigma} $ derivatives, since the good factors that have been prepared
by the slice testing expansion are more than enough to pay for these uncompensated counterterms.

Hence we can write informally the Bosonic integrand after action of derivatives as
\bea\label{eq:bosogauss}
&& \int d \nu_\cB  F_\cB = \int d \nu_\cB e^{ V(\cB, J_a)}  \sum_{G_\cB \in \cG (\cB)} A_{G_\cB} (\sigma)
 \; .
\eea 
where the graphs in $\cG (\cB) $ are   resolvent graphs, but which are no longer minimal, nor connected
over all vertices of $\cB$. They have still the same set of marks $J(\cB)$, but their order in $\lambda$ is now
bounded by $\vert J( \cB) \vert + \vert \cB -1 \vert $, since the tree of $\vert \cB \vert -1$ derivatives connecting
$\cB$ brings down exactly $\vert \cB \vert -1$ factors $\lambda$ from the exponential and from resolvents. 

Furthermore the amplitudes $ A_{G_\cB} (\sigma)$ are given by a formula intermediate between 
\eqref{ampres} and \eqref{ampresren}.
There can be up to $2\vert \cB \vert -1$ additional (uncompensated) tadpole insertions. Furthermore the 
resolvent of the additional tadpoles may not be $R-1$ but can be $R$'s. 

Then we perform a Cauchy-Schwarz inequality with respect to the positive measure $d\nu_\cB$ to separate the graphs from the 
remaining interaction:
\bee  \label{CS}   \sum_{G_\cB \in \cG (\cB)}   \Bigl{|}  \int d \nu_\cB e^{ V(\cB, J_a)}  (\sigma) A_{G_\cB} (\sigma) \Bigl{|} \;  \le \; 
 \sum_{G_\cB \in \cG (\cB)}   \Bigr{(} \int d \nu_\cB  \prod_a  e^{2 \vert  V(\cB, J_a) \vert }   \Bigr{)}^{1/2}  
\Bigr{(} \int d \nu_\cB   \vert  A_{G_\cB} (\sigma)  \vert^2  \Bigr{)}^{1/2} .
\ee

\subsection{Non-Perturbative Bound I: The Remaining Interaction}
In this subsection we bound $ \int d \nu_\cB  \prod_a  e^{2 \vert V(\cB, J_a) \vert } $ in \eqref{CS}.

\begin{lemma}\label{boundv}
For $g$ in the cardioid domain $\cC  ard_\rho$ defined in Theorem \ref{thetheorem}
we have
\bea \vert \exp( {\rm Tr}  V_j (\sigma) )  \vert &\le&  \exp \bigl( 0(1)\,   \rho \,  j +  \vert \lambda \vert^{1/2}  \sin ( \phi/2)     {\rm Tr} \, \sigma  + \rho {\rm Tr}  \bigl(  C_{\le j} \sigma C_j   \sigma \bigr)] \bigr).
\label{boundvj}
\eea
\end{lemma}
\prf   
Using \eqref{slicedinter} we write, putting ${\rm Arg}\; \lambda = \phi $
\bea
\vert \exp( {\rm Tr}   V_j (\sigma) )  \vert  & = & \exp \Re \bigl( 
\int_0^1  du_j \Tr  \bigl[ 6 \lambda T_{j} T_{\le j} (t )  + 2i \sqrt{\lambda} T_{j} \sigma 
 - i \sqrt{\lambda}  [ R_{\le j}(t) -1 ]  C_{j} \sigma \bigr]  \vert \bigr)
 \nonumber \\
&\le & \exp \bigl(  0(1)\;  \bigl( \rho  j +  \vert \lambda \vert^{1/2}\sin ( \phi/2)   {\rm Tr}  \, \sigma  + \vert \lambda \vert
 \vert{\rm Tr} \bigl(   C_j \sigma  R_{\le j} C_{\le j}  \sigma   \bigr)\vert \bigr)\bigr)
\nonumber \\
&\le& \exp \bigl(  0(1)\;  \bigl( \rho  j +  \vert \lambda \vert^{1/2}\sin ( \phi/2)    {\rm Tr} \, \sigma  + \rho\;    {\rm Tr}  \bigl(  C_{\le j}^{1/2}  \sigma C_j   \sigma C_{\le j}^{1/2}  \bigr) \bigr)
\nonumber \\
&=& \exp \bigl(  0(1)\;  \bigl( \rho  j +  \vert \lambda \vert^{1/2}\sin ( \phi/2)     {\rm Tr}\, \sigma  + \rho \;    {\rm Tr}  \bigl(  C_{\le j}  \sigma C_j   \sigma  \bigr)\bigr).  \label{noper1} 
\eea
For the first to second line we used that $ ( R_{\le j}(t) -1 )  C_{j} \sigma  =   -2i \sqrt{\lambda}  R_{\le j} C_{\le j}  \sigma  C_j \sigma $.
Then for $A$ positive\footnote{We usually simply say positive for "non-negative", i. e. each eigenvalue is strictly 
positive or zero.} Hermitian and $B$ bounded we have $\vert  \Tr A B \vert \le \Vert B \Vert \Tr A $. 
Indeed if $B$ is diagonalizable with eigenvalues $\mu_i$, computing the trace in a 
diagonalizing basis we have $\vert \sum_i   A_{ii} \mu_i \vert \le \max_i \vert \mu_i \vert \sum_i   A_{ii}  $;
if $B$ is not diagonalizable we can use a limit argument. 
We can now remark that for any Hermitian operator $L$ we have, if $\vert {\rm Arg}\; \lambda \vert = \vert \phi\vert  < \pi$, 
$\Vert ( 1- i \sqrt \lambda L )^{-1}\Vert \le \frac{1}{\cos (\phi /2)}$. We can therefore
apply these arguments to $A = C_{\le j}^{1/2}  \sigma C_j    \sigma C_{\le j}^{1/2} $ (which is Hermitian positive)
and $B = C_{\le j}^{-1/2} R_{\le j} C_{\le j}^{1/2}$.
Indeed 
\bee \Vert B \Vert =  \Vert R_{\le j} \Vert  = \Vert ( 1- i \sqrt  \lambda C_{\le j}^{1/2} \sigma C_{\le j}^{1/2} )^{-1}\Vert \le  \frac{1}{\cos (\phi /2)}. 
\label{resobound}
\ee
We conclude since in the cardioid $ \frac{\vert \lambda \vert}{\cos^2 (\phi /2)} \le \rho$.  \qed

\medskip
We can now bound the first factor in the Cauchy-Schwarz inequality \eqref{CS}.
\begin{theorem}[Bosonic Integration]\label{BosonicIntegration}
For $\rho $ small enough and for any value of the $w$ interpolating parameters
\bea  \Bigl{(} \int d\nu_\cB e^{ 2 \sum_{j \in J(\cB)} \vert V_{j} (\sigma_a) \vert }   \Bigr{)}^{1/2}  &\le& 
e^{ O(1) \rho  \sum_{j \in J(\cB)} j }.  \label{rhobound}
\eea
\end{theorem}
\prf  
Remark that the first term in $ 0(1)\, \rho\,  j $ in \eqref{boundvj} gives precisely a bound in $O(1) \rho  \sum_{j \in J(\cB)} j $. So it remains to check that
\bee\int d\nu_\cB e^{ O(1) \bigl(\vert \lambda \vert^{1/2}\sin ( \phi/2)     {\rm Tr}\, \sigma  + \rho \;    {\rm Tr}  (  C_{\le j}  \sigma C_j   \sigma  ) \bigr)}   \Bigr{)}^{1/2}  \le 
e^{ O(1) \rho  \sum_{j \in J(\cB)} j }.
\ee

Applying Lemma \ref{boundv} we get
\bee
 \int d\nu_\cB  \prod_{a \in \cB}   e^{ 2 \sum_{j \in J_a} \vert V_{j} (\sigma_a) \vert }  \le 
\int d\nu_\cB  \; e^{ \; \frac{1}{2} < \sigma , \bQ \sigma >  + < \sigma, \bP >}
\ee
where $\bQ$ is a symmetric positive matrix in the big vector space  $\bV$ which is the tensor product of the spatial space $L_2 ([0,1]^2) $
with the ``replica space" generated by the orthonormal basis $\{e_a\}$, $a \in \cB$.
$\bP$ is a constant function in $L_2 ([0,1]^2) $, which takes the (single) value $ P=O(1) \vert \lambda \vert^{1/2}\sin ( \phi/2) $.

More precisely $\bQ$ is diagonal in replica space and $\bQ$ and $\bP$ are defined by the equations
\bea  < \sigma , \bQ \sigma >  &=& \sum_{a \in \cB} \sum_{j \in J_a} < \sigma_a , Q_j \sigma_a >, \  < \sigma_a , Q_j \sigma_a > \equiv
 O(1)  \rho  {\rm \ Tr}  \bigl(  C_{\le j}   \sigma_a C_{j}  
 \sigma_a \bigr), \label{niceequ1}\nonumber\\    
 < \sigma , \bP> &=&  \sum_{a \in \cB}  \sum_{j \in J_a}  P  \Tr \sigma_a = P \sum_{a \in \cB} \vert J_a \vert \Tr \sigma_a
 \label{niceequ2}.
\eea
Each $Q_j$ is positive and using the bounds \eqref{multbound} it is easy to check that the kernel of $Q_j$ is bounded by
\bee \label{boundqj}
Q_j (x,y) \le O(1) \rho\, j  e^{- M^{j} \vert x-y \vert }.
\ee
Hence, since we work on a fixed square of unit volume in ${\mathbb R}^2$ the following lemma follows easily.
\begin{lemma}
Uniformly in $j_{max}$
\bea
{\rm Tr}\; Q_j &\le&  O(1)  \rho  j ,\label{bonnetrace}\\
\Vert Q_j \Vert  &\le& O(1) \rho  M^{-j/2} \label{bonnenorm}.
\eea
\end{lemma}
The second bound \eqref{bonnenorm} is absolutely not optimal but enough for what is needed below. 
It just comes from bounding first the norm of $e^{- M^{j} \vert x-y \vert }$ by its Hilbert-Schmidt norm
which is proportional to $M^{-j}$ (we are in two dimensions) and then bounding the $j$ factor in \eqref{boundqj} by $M^{j/2}$.

The covariance $\bX$ of the Gaussian measure $d\nu_\cB$ is a symmetric matrix on the big space $\bV$, which is the tensor product 
of the identity in space times the matrix $X_{ab} (w_{\ell_B} )$ in the replica space. 
Defining $\bA \equiv \bX \bQ$, we have
\begin{lemma}
The following bounds hold uniformly in $j_{max}$ 
\bea
\Tr \; \bA &\le&  O(1)  \rho  \,  \sum_{j \in J(\cB)} j  ,\label{bonnegtrace}\\
\Vert \bA \Vert  &\le& O(1) \rho  \label{bonnegnorm}.
\eea
\end{lemma}
\prf Since $\bQ = \sum_{a \in \cB} \sum_{j \in J_a} \bQ_j $ is diagonal in replica space we find that 
\bee
\Tr \; \bA = \sum_{a \in \cB} \sum_{j \in J_a} \Tr \bX \bQ_j = \sum_{a \in \cB} \sum_{j \in J_a}  X_{aa} (w_{\ell_B} ) \Tr  Q_j = \sum_{a \in \cB} \sum_{j \in J_a}{\rm Tr}\; Q_j \le  O(1)  \rho  \,  \sum_{j \in J(\cB)} j .
\ee
where in the last inequality we used \eqref{bonnetrace}. Furthermore by the triangular inequality in \eqref{niceequ1} and using \eqref{bonnenorm}
\bea
\Vert \bA \Vert  \le   \sum_{a \in \cB}  X_{aa} (w_{\ell_B} )  \sum_{j \in J_a} \Vert  Q_j \Vert = \sum_{a \in \cB} \sum_{j \in J_a}  O(1)  \rho  M^{-j/2} \le O(1)   \rho.
\eea
where we used the fundamental fact that all vertices $a\in\cB$ have \emph{disjoint subsets of scales} $J_a$.
\qed

We can now complete the proof of Theorem \ref{BosonicIntegration}. 
Since $  \Tr \bA^n  \le  \Tr \bA   \Vert \bA \Vert^n $, by \eqref{bonnegnorm} for $\rho$ small enough the series $\sum_{n=1}^\infty \Tr \bA^n $ converges and is bounded by $ 2\Tr \bA$. This justifies the computation
\bea \int d\nu_\cB  \; e^{\frac{1}{2}  <\sigma, \bQ \sigma > + <\sigma ,\bP > }  &=& e^{\frac{1}{2} <\bP , \bX (1-\bA)^{-1}  \bP > } [ \det(  1-\bA)  ]^{-1/2}  
\nonumber  \\ \,
[ \, \det (1 - \bA) ]^{-1/2} 
&=& e^{\frac{1}{2}  \sum_{n=1}^\infty (\Tr \bA^n)/n  }   \le e^{\Tr \bA } \le e^{ O(1)  \rho  \sum_{j \in J(\cB)}  j  }  .
\eea
Moreover
\bea
e^{\frac{1}{2} <\bP , \bX (1-\bA)^{-1}  \bP > }  &\le&    e^{    \frac{1}{2} \Vert (1-\bA)^{-1}  \Vert <\bP , \bX \bP > }   \le   e^{ O(1)  \rho   \sum_{ab} \vert J_a \vert   X_{ab} (w_{\ell_B} )  \vert J_b \vert  }
\nonumber
\\  &\le&   e^{ O(1)  \rho   \sum_{a \in \cB , b\in \cB} \vert J_a \vert   . \vert J_b \vert  } =  e^{ O(1)  \rho  \vert J(\cB) \vert^2 } 
\le e^{ O(1)  \rho  \sum_{j \in J(\cB)}  j  }  . 
\label{goodtra}
\eea
where we used that $  X_{ab} (w_{\ell_B} ) \le 1$ for any $\{w_{\ell_B}\}$ and for the last inequality we used again that vertices $a\in\cB$ have disjoint subsets of scales $J_a$.

This completes the proof  of Theorem \ref{BosonicIntegration}. \qed

\subsection{Non-Perturbative Bounds II: Getting Rid of Resolvents}
In this subsection we now explain how to bound, for a fixed $G_\cB$, the second factor 
\bee I= \bigl{(}\int d \nu_\cB(\sigma )   \vert  A_{G_\cB} (\sigma)  \vert^2   \bigr{)}^{1/2} 
\ee
of the Cauchy-Schwarz inequality \eqref{CS}. We shall not try to establish sharp bounds on $I$, just bounds sufficient for the proof of Theorem \ref{thetheorem}.
This is still a non-perturbative problem, since the resolvents in the $A_{G_\cB} (\sigma)$, if expanded in power series of $\sigma$
and integrated out with respect to $ d \nu_\cB  $, would lead to infinite divergent series of Feynman graphs. Hence we shall
use the norm bound  \eqref{resobound} to get rid of these resolvents.

For this we write first
\bee I^2 = \int d \nu_\cB   A_{G_\cB} (\sigma)  A_{\bar G_\cB} (\sigma)  
\ee
where $\bar G_\cB (\sigma) $ is the complex conjugate graph, made with complex conjugate resolvents.
We consider $G_\cB \cup \bar G_\cB$ as a (not connected) graph with twice the number of propagators and vertices of $G_\cB$. 

Let us call $n\ge 2$ the order of perturbation theory for $G_\cB $ (i.e. the number of its $\sigma$ propagators).
 The order of perturbation for $G_\cB \cup \bar G_\cB$ is $2n$, and the number of its c-propagators
is $4n$. Ordering the slices of $J(\cB)$ as $j_1 < j_2 \cdots < j_p$, we
have $p=\vert J(\cB) \vert \ge n/2 $ marked c-propagators, one for each slice $j_k$, $k=1, \cdots, p$. 
Indeed $G_\cB$ came from a minimal resolvent graph $G \in \cG$ for a certain set of slices $J(\cB)$,
and the definition of minimality implied that the order of $G$ was at most $\vert J(\cB)\vert $; moreover $G_\cB$ was obtained from $G$ by adding at most $\vert \cB\vert -1$ further $\sigma$ propagators through the MLVE (see subsection \ref{bosointegr}).

We need now to better explicit the fact that the marked c-propagators cannot be tadpoles. This is 
indeed the source of the good factors which make the expansion converge.
For this we simply expand the $R-1$ factors of the tadpoles of the graph $G$ as $\pm 2i \sqrt \lambda R C \sigma $.
Then we contract all the $\sigma$ fields produced in this way. 
Each Wick contraction step decreases the number of these fields by two if they contract between themselves and by one if they contract to $R$ factors,
hence in any such Wick contraction the number of such $\sigma'$s decreases at least by one. At the end of this step we obtain a set of new 
resolvent graphs which have the following properties:

\begin{itemize}
\item they have a set $M$ of marked c-propagators which are now explicitly not tadpoles, with indices $j_1 < j_2 \cdots < j_p$,
with $p \ge n/2$; moreover all these marked propagators have value $C_{j_k}$, not $R C_{j_k} $. We say that they have been cleaned from their
resolvent factors.

\item their total order of perturbation theory (number of $\sigma$-propagators) is at most $4n$.

\item the not marked c-propagators have factors either $C$ or $RC$.

\end{itemize}

The amplitudes for these graphs are given by \eqref{ampres} hence still include a resolvent $R$ for typically many c-propagators.

Our next step is to get rid of all these $R$ factors, essentially using the fact that their norm is bounded by $\cos^{-1} (\phi /2)$ in the cardioid domain.
This is not trivial and we shall rely on the technique of recursive Cauchy-Schwarz (CS) inequalities of \cite{sefu1,Delepouve:2014bma}. 
Consider a fixed connected intermediate field graph $G$ which is a connected component of the previous list, with order $m \le 4n$. 
It has $2m$ c-propagators. The key definition to define the CS inequality  is that of a balanced cut. Since it is recursive, we need first to consider a 
slightly more general class of resolvent amplitudes, for intermediate field vacuum graphs $G$ which have a particular 
set $R$ of their c-propagators  called resolvent propagators, since they bear resolvents. The other c-propagators, bearing no resolvents, are called \emph{cleaned}.
The amplitudes of such partly cleaned graphs are given by the following formula which generalizes \eqref{ampres}
\bee  A_{G,R} (t, \sigma) =  \prod_{v \in V(G)} \bigl[ (-\lambda) \int_{[0,1]^2}  d^2 x_v  \bigr]  \prod_{\ell \in R} [ R(\sigma)C_{j(\ell)}  ] (x_\ell, x'_\ell )
\prod_{\ell \not\in R} [ C_{j(\ell)} ] (x_\ell, x'_\ell ).
\label{amppartres}
\ee 
Hence $R = CP(G)$, the full set of $c$-propagators of $G$, correspond to the resolvent amplitudes \eqref{ampres}, and $R=\emptyset$ correspond
to the ordinary perturbative Feynman amplitudes for intermediate field graphs. 
This 
definition generalizes in a straightforward way to the case where c-propagators have
certain slice restrictions.

\begin{figure}[!t]
\begin{center}
{\includegraphics[width=9cm]{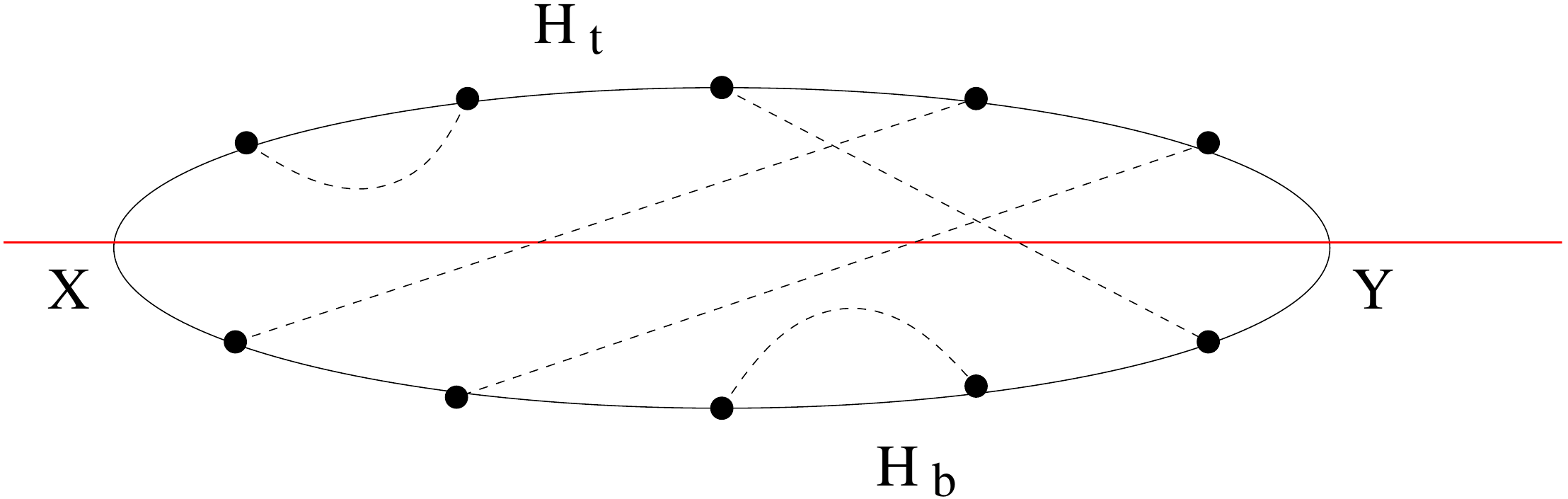}}
\end{center}
\caption{A resolvent graph with a balanced cut. Here $\sigma$-propagators are pictured as dotted lines.}
\label{cs1}
\end{figure}

We define a \emph{balanced $X-Y$ cut} for $(G,R)$ as a partition of the graph $G$ into two pieces
which we call the top and bottom chains, $H_t$
and $H_b$, each containing the \emph{same number of resolvent propagators} (up to one unit if the initial number of resolvent propagators is odd).
 $H_t$ and $H_b$ are each made of a chain of $p$ or $p-1$ resolvent propagators, 
plus the two half resolvent propagators $X$ and $Y$ at the ends of the chain, plus an arbitrary number of cleaned propagators (it needs not be the same number 
in $H_t$ and $H_b$). To these two chains are hooked the same number $q$
of half $\sigma$-propagators which cross the cut, plus 
inner $\sigma$-propagators which do not cross the cut. 
Remark that $\sigma$ propagators have no reason to occur at 
symmetric positions along the top and bottom chains (see Figure \ref{cs1}).

Balanced cuts for an intermediate field connected vacuum graph $(G,R)$ with $R \not = \emptyset$ can be obtained in many different ways. 
A nice way to define such cuts is to first select a spanning tree of $\sigma$ propagators of $G$. Then turning around the tree provides a well defined cyclic ordering
of the $2m$ c-propagators of the graph (jumping over the $\sigma$-propagators not in the tree). Balanced
cuts are then obtained by first \emph{contracting all cleaned propagators along the cycle}, and then selecting an antipodal\footnote{Or almost antipodal if the number of resolvents is not even; this can happen only at the first
CS step.} pair $(X, Y)$
among the resolvent propagators left in that cycle. We then
cut the cycle across that pair (see Figure \ref{cs1}).

\begin{figure}[!t]
\begin{center}
{\includegraphics[width=7cm]{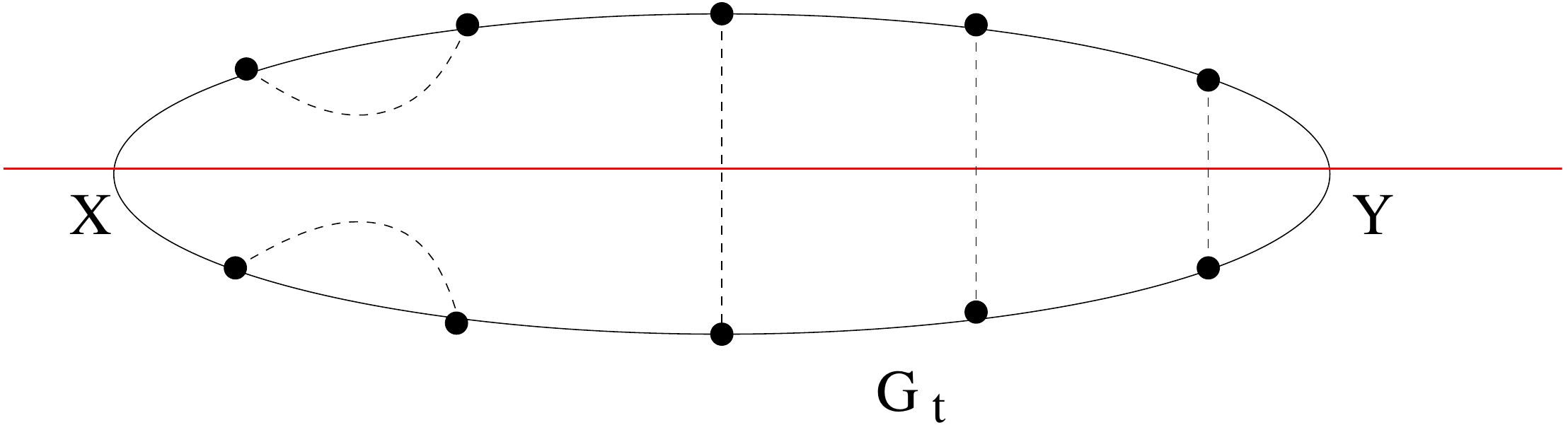} \hskip.5cm\includegraphics[width=7cm]{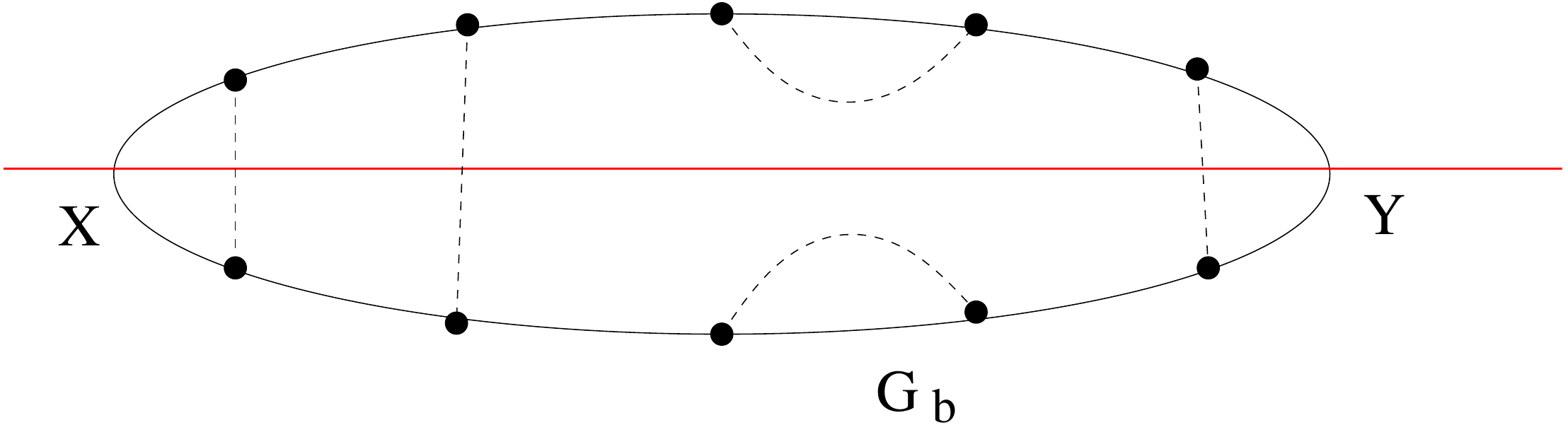}}
\end{center}
\caption{The top and bottom mirror graphs for the balanced cut}
\label{cs2}
\end{figure}

To any such balanced cut is associated a Cauchy-Schwarz (CS) inequality. It bounds the
resolvent amplitude $A_G$ (see \eqref{ampres}) by the geometric mean of the amplitudes of the two graphs
$G_t = H_t \cup \bar H_t$ and $G_b = H_b \cup \bar H_b$. These two graphs are
obtained by gluing $H_t$ and $H_b$ with their mirror image along the cut. Remark that in this gluing the $\sigma$ propagators crossing the cut are 
fully disentangled: in $G_t$ and $G_b$ they no longer cross each other, see Figure \ref{cs2}. Remark also
that the right hand side of the CS inequality, hence the bound obtained for $A_G$, is a priori different for different balanced cuts.

In such a CS inequality, something crucial can be gained. In the cardioid we recall that by \eqref{resobound} $\Vert R\Vert \le \cos^{-1} (\phi/2) $.

Now the two propagators $X$ and $Y$ crossed by the balanced cut
at the end of the top and bottom chain, which had values $RC$ in the amplitude $A_G$,  can be replaced by two \emph{ordinary} 
propagators $C$ in the amplitudes of $G_t$ and $G_b$, loosing simply a factor $\cos^{-2} (\phi/2)$ for the two norms of $R$.
Hence they are cleaned.

\begin{lemma} \label{CSstep}
For any balanced $X-Y$ cut
\bee  \vert A_{G,R} (\sigma) \vert   \le \frac{1}{\cos^2 (\phi /2)} \sqrt{A_{G_t, R - \{X \cup Y\}}(\sigma)}\sqrt{A_{G_b, R - \{X \cup Y\}}(\sigma)}
\ee
uniformly in $\sigma$.
Hence we have cleaned the two resolvent propagators $X$ and $Y$ crossed by the cut.
\end{lemma}
\prf
Each $CS$ inequality is simply obtained by writing
\bee <H_t , \cO H_b  > \le \Vert \cO \Vert  \sqrt{<H_t H_t >} \sqrt{<H_b H_b >} \label{tensocs}
\ee
in the tensor product of $2+q$ Hilbert spaces corresponding to the two end c-propagators and the $q$ crossing $\sigma$-propagators. 
We symmetrize first the operators $RC$ of the two cut propagators, writing them as $C^{1/2}B C^{1/2} $  with $ B = C^{-1/2}R C^{1/2} $.
The operator $\cO = B \otimes \Pi \otimes B$ 
in  \eqref{tensocs} is the tensor product of the two end operators $B$ and of a permutation operator $\Pi$ for the remaining
$H^{\otimes q}$ tensor product of the $q$ crossing $\sigma$-propagators. Therefore
$\Vert \cO \le \Vert B \Vert ^2  \Vert \Pi \Vert $. Any permutation operators has eigenvalues which are roots of unity, hence has norm bounded by 1, and 
$\Vert  B \Vert = \Vert R \Vert \le \cos^{-1} (\phi/2)$. 
\qed

Again remark that this lemma can also be applied to the case where the c-propagators have
certain slice restrictions, in which case these restrictions are carried to their copies in $G_t$ and $G_b$.

Starting with a full resolvent graph, the inductive CS inequalities of \cite{sefu1,Delepouve:2014bma} consist in iterating lemma 
\ref{CSstep} until no resolvents are left anywhere. In this way we can therefore reach a bound 
made of a geometric mean of $2^m$ ordinary perturbative amplitudes for an initial resolvent graph of of order $m$. 
To understand the result of the induction, let us observe that
\begin{itemize}

\item Only at the first step the number of resolvent propagators can be odd. In that case 
we choose an almost antipodal pair (antipodal up to half a unit): 
but at all later stages the mirror gluing creates an even number of resolvents and we can choose truly antipodal pairs. 

\item The result of $m$ complete inductive layers of CS steps applied to a starting graph $G$ of order $m$ is a family $F_m^\cC(G)$ of $2^m$ graphs,
which depends on the inductive choices of all the balanced cuts of the induction. The set of these choices is noted $\cC$.
The graphs of $F_q^\cC(G)$ are called $q$-th layer graphs and can be pictured to stand at the leaves of a rooted binary tree, with the initial graph
$G$ standing at the root. $\cC$
is a choice of a balanced cut for each vertex of that rooted binary tree. It splits the parent graph $G'$ of layer $q-1$ for the parent
edge of the vertex into into two children graphs $G'_t$ and $G'_b$ of layer $q$, one for each of the two children-edges of the vertex.

\item Although the graphs in the family $F_q^\cC(G)$ may have very different orders, they all have the same number of resolvents (up to one at most, if the initial
number of resolvents was odd).

\item No matter which inductive choice $\cC$ is made, every 
c-propagator $\ell$ of the initial graph $G$ gets finally copied into exactly $2^m$  c-propagators in the union of all graphs
of $F_m^\cC(G)$. Notice that all these copies have the same slice-attribution $j(\ell)$ than the initial propagator.
But they are not at all evenly distributed among the members of the family,

\end{itemize}

This is summarized in the following lemma.

\begin{lemma}
For any choice $\cC$ of $m$ recursive cuts
\bee  \vert A_{G} (\sigma) \vert   \le  \bigl[ \prod_{G' \in F_n^\cC(G)} \vert A_{G'}  \vert \bigr]^{2^{-m}}
\ee
uniformly in $\sigma$. The amplitudes $A_{G'}$ are computed with coupling constants $\rho$ instead of $\vert \lambda \vert$.
\end{lemma}
\prf
Straightforward induction using lemma \ref{CSstep}. We bound all the factors  $\cos^{-2} (\phi/2)$ generated by the CS inequalities by changing the factor $\vert \lambda^{V(G')} \vert$ 
into $\vert \rho^{V(G')} \vert$. Indeed for each pair of resolvents destroyed by a CS inequality there is an independent coupling constant factor $\vert \lambda\vert$,
and in the cardioid we have $\vert \lambda\vert\cos^{-2} (\phi/2) \le \rho$.
\qed

Recall that the amplitudes $A_{G'}$ have no resolvent factors any more, hence are ordinary perturbative amplitudes
no longer depending on $\sigma$. We can now reap the good factors due to the marked propagators and use them to bound all
remaining amplitude factors and combinatorics.

\subsection{Perturbative Bounds, Combinatorics and Final Bound}
\label{finalsubsec}

We shall be brief, as this section does not contain any new idea. Summarizing the results
of the previous section we obtained a geometric average over amplitudes for 
ordinary graphs $G'$ (not necessarily connected) each with order at most $4n$, and at least $n/2$ different marked slice propagators 
$j_1 < j_2 < \cdots < j_p$, each of which is not a tadpole. The sum over slices for all other propagators of the graph are restricted to  $J(\cB) = \{j_1, \cdots j_p\}$
(since all slices with indices not in $J(\cB)$ were deleted when their parameter $t_j$ was put to 0 in \eqref{testing1}).

\begin{lemma} The amplitude for any such graph $G$ is bounded by
\bee \vert A_{G'} \vert \le O(1)^n \rho^n p^{8n} M^{- 2\sum_{i=1}^p  j_i /7}   \label{pertubound}
\ee
\end{lemma}

\prf We bound the propagators according to \eqref{multbound}. Consider the highest marked propagator, of scale $j_p$. Since it is not a tadpole we gain an integration factor $M^{-2j}$ for 
the integration of one of the two vertices hooked to it with respect to the other. Then we cross the (at most 6) propagators 
which touch that propagator and consider the next highest uncrossed marked propagator. Iterating,
we gain a factor at least $M^{- 2\sum_{i=1}^p  j_i /7} $ (with little extra care we could have improved this bound to $M^{- \sum_{i=1}^p  j_i /2} $).
All other vertex integrations can be bounded by $1$ and each sum over
slice attributions can be bounded by $p$. Since there are at most $8n$ such sums we obtain \eqref {pertubound}.
\qed

Taking $\rho $ small enough, with a fraction of the factor $M^{- 2\sum_{i=1}^p  j_i /7} $ in 
\eqref{pertubound} we can bound the factor  $e^{ O(1) \rho  \sum_{j \in J(\cB)} j }$ in 
\eqref{rhobound}.

The sum over all combinatorial structures of Feynman graphs at order $n$ and
the sum over the forests of section \ref{multilve} are all similarly bounded by $O(1)^n n^{4n}$. We refer for
more details on exact values for such combinatorial factors to \cite{MLVE}.
Since $p \ge n/2$ and our graphs have order at most $4n$ we have 
\bee  O(1)^n n^{4n} M^{- O(1) p^2}\le    [ O(1)  \rho]^n .
\ee
Finally let us describe how to bound the additional factors due to the auxiliary expansion

\begin{itemize}

\item  the sum over partitions $\Pi$ is the Bell number $B_p$, itself bounded by $p!$

\item  the sum over trees $\cT$ is bounded by $p^{p-2}$ by Cayley's theorem

\item  the $w$ integrals and factors are all bounded by 1

\item  the crossed counterterms or tadpoles are bound by $p^{p-1}$

\item the sum over derivatives actions $\partial w$, hence over the modified graphs is certainly bounded by $p^{4p}$

\end{itemize}

In conclusion all auxiliary expansion effects factorize over the blocks and in each block of size $p$ are bounded by $p^{O(1)p}$,
hence easily beaten by another small fraction of the $M^{- O(1) p^2}$ convergence factor due to the marked renormalized propagators.

Hence we obtain a uniformly convergent bound for the series \eqref{treerep} in the cardioid. 
This achieves the proof of Theorem \ref{thetheorem}.

\medskip
\noindent{\bf Acknowledgments}  We thank T. Delepouve and R. Gurau for useful discussions.

\end{document}